\documentclass[pra,a4paper,twocolumn,superscriptaddress,longbibliography,nofootinbib]{revtex4-2}

\usepackage{amssymb}
\usepackage{amsmath}
\usepackage{amsfonts}
\usepackage{graphicx}
\usepackage{bm}
\usepackage[dvipsnames]{xcolor}
\usepackage{multirow}
\usepackage{comment}
\usepackage{subfigure} 
\usepackage[colorlinks]{hyperref}

\DeclareMathOperator{\sgn}{sgn}

\newcommand{\intl}{\int\limits}

\begin{document}

\title{Deformation of a N\'eel-type Skyrmion in a Weak Inhomogeneous Magnetic Field: Magnetization \textit{Ansatz} and Interaction with a Pearl Vortex}

\author{S. S. Apostoloff}

\affiliation{\hbox{L. D. Landau Institute for Theoretical Physics, Semenova 1-a, 142432, Chernogolovka, Russia}}
\affiliation{Laboratory for Condensed Matter Physics, HSE University, 101000 Moscow, Russia}

\author{E. S. Andriyakhina}

\affiliation{\hbox{L. D. Landau Institute for Theoretical Physics, Semenova 1-a, 142432, Chernogolovka, Russia}}

\author{I. S. Burmistrov}

\affiliation{\hbox{L. D. Landau Institute for Theoretical Physics, Semenova 1-a, 142432, Chernogolovka, Russia}}
\affiliation{Laboratory for Condensed Matter Physics, HSE University, 101000 Moscow, Russia}

\date{\today} %, v8}

\begin{abstract}
In this work, we develop a theory of (meta)stable states of N\'eel-type skyrmions in weak nonuniform magnetic fields. We claim an ansatz for modeling the non-symmetric magnetization that can be implied for both analytics and numerical simulations. Our theory accounts for changes in the size of skyrmion parameters and also includes deformations from the centrally symmetric shape. The \textit{Ansatz} streamlines the analytic calculation of the skyrmion free energy, enhancing the efficiency of the minimizing process.
Performing the minimization in two stages, one can find all the minima, global and local, of the free energy, discovering the stable and metastable states. 
We apply the developed methodology to investigate the (meta)stable configurations of skyrmions influenced by the stray field of a Pearl vortex. Our study reveals the dependence of skyrmion spatial parameters on the vortex field effective strength and presents a phase diagram identifying regions where metastable configurations are predicted. Corroborated by micromagnetic simulations, our findings offer a detailed perspective on the interaction between magnetic skyrmions and superconducting vortices.
\end{abstract}

\maketitle

\section{Introduction}

Magnetic skyrmions, predicted in the late 1980s~\cite{Bogdanov1989} and discovered in recent decades \cite{Muhlbauer2009,Yu2011,Seki2012}, have rapidly ascended to prominence in the field of magnetism and novel materials, offering new paradigms for information storage and processing. They are distinguished by their topological stability~\cite{Bogdanov1994,Nagaosa2013}, which renders them resistant to perturbations, and their small size, which makes them ideal candidates for high-density storage applications. The interplay between skyrmions and magnetic fields is a cornerstone of skyrmionics, influencing their formation, stability, and dynamic properties~\cite{Muhlbauer2009,Yu2011,Seki2012,Yu2010,Jiang2015,Herve2018,CortesOrtuno2019}. This relationship is particularly nuanced in the presence of non-uniform fields, where skyrmions exhibit a range of behaviors that are critical to their practical utility~\cite{CarvalhoSantos2015,Wang2019,Shustin2023}.

The study of superconductor–chiral ferromagnet (SF) bilayers has become increasingly relevant in this context. Over the past few decades, the intricate relationship between magnetism and superconductivity in heterostructures has piqued interest among researchers~\cite{Ryazanov2004,Lyuksyutov2005,Buzdin2005,Bergeret2005,Eschrig2015}. These bilayers are particularly intriguing as they can host both magnetic skyrmions, which emerge through the effects of the Dzyaloshinskii–Moriya interaction~\cite{Bogdanov1989}, and superconducting vortices~\cite{ABRIKOSOV1957199, Pearl1964}. The discovery of skyrmions in SF bilayers has led to intensive investigations into their physical attributes and prospective technological applications. Skyrmions have been shown to give rise to bound states similar to Yu-Shiba-Rusinov states~\cite{Pershoguba2016,Poyhonen2016} and to modify the Josephson effect between superconductors~\cite{Yokoyama2015}, as well as to affect the superconducting critical temperature~\cite{Proshin2022}.

The coexistence of skyrmions and superconducting vortices introduces intriguing dynamics. They are known to form bound pairs influenced by various mechanisms, including spin-orbit coupling, proximity effects, and magnetic stray fields~\cite{Hals2016,Baumard2019,Dahir2019,Menezes2019,Dahir2020,Andriyakhina2021,Andriyakhina2022,Apostoloff2023}. Recent works~\cite{Chen2015, Yang2016, Gungordu2018, Mascot2019, Rex2019, Garnier2019, Rex2020, Gungordu2022, Nothhelfer2022, Konakanchi2023} suggest that these pairs might offer a foundation for topological quantum computing.

From an experimental perspective, the interplay between skyrmions and superconductivity has seen recent advancements. Observations have shown that skyrmions can modulate vortex dynamics~\cite{Palermo2020}, couple with vortices in chiral magnet-superconductor heterostructures~\cite{Petrovic2021}, and even slightly enlarge due to the presence of Pearl vortices~\cite{Machain2021}. Importantly, their potential in next-generation computing and memory devices is underscored by the capability to precisely manipulate single skyrmions using tunneling microscopes~\cite{Romming2013}.

In our research, we focus on the interactions between a skyrmion and a vortex via stray fields. We propose that a thin insulator layer separates the ferromagnetic and superconducting layers, mitigating the quantum proximity effect. A sketch of the considered setup can be found in Fig.~1 of Ref.~\cite{Andriyakhina2021}. Commonly, studies have overlooked the subtle changes in skyrmion structure due to stray fields. However, as highlighted in Ref.~\cite{Apostoloff2023}, these changes play a significant role in the overall understanding of the system. In the case of coaxial configurations, interaction of a skyrmion with a vortex can lead to significant changes in the size of the skyrmion, its shape, and even its chirality \cite{Apostoloff2023}.

Recent theoretical works~\cite{Andriyakhina2021,ru-JETP-letters-2022} have shown that stable bound states of a skyrmion and a vortex can exist in both coaxial (with $a=0$, where $a$ represents the relative distance between the centers of the skyrmion and the vortex) and eccentric (with $a\neq 0$) configurations. Solving the exact Euler-Lagrange equations for the minimization problem of the total free energy, presented in Eq.~\eqref{eq:F_tot}, is a notoriously difficult problem. Below, we propose a variational method based on
an effective approximation to the exact solution, circumventing the need to solve the exact Euler-Lagrange equation directly.

In this study, we explore the effects of weak magnetic fields on skyrmion behavior in thin films, particularly focusing on N\'eel-type skyrmions. Our \textit{Ansatz}, as detailed in Eq. \eqref{eq:ansatz_eccentric} and illustrated in Figure~\ref{fig:ansatz}, describes skyrmion magnetization in arbitrary weak inhomogeneous magnetic fields. These fields are considered weak enough to only slightly disrupt the cylindrical symmetry of a skyrmion. The \textit{Ansatz} is pivotal for understanding skyrmion deformation. %, especially in the context of their interaction with a superconducting Pearl vortex in thin films. 
It effectively predicts the skyrmion's stable position and spatial parameters, such as radius and domain wall width, and identifies metastable states beyond the reach of conventional micromagnetic simulations.

The \textit{Ansatz}'s main concept involves decomposing skyrmion magnetization into two parts: a cylindrically symmetric component resulting from the magnetic field averaged over the polar angle around the skyrmion center, and a component accounting for deviations in cylindrical symmetry due to the actual magnetic field distribution. While our focus is on Neel-type skyrmions, we suggest that the fundamental approach of the \textit{Ansatz} could be applicable to Bloch-type skyrmions, though this requires further verification.

For clarity, we define some key notations: we assume that a skyrmion exists within a spatially inhomogeneous magnetic field $\bm{B}(\bm{r})$. A coaxial configuration implies that the skyrmion's center is at the point $\bm{a} = 0$, particularly relevant as we later consider the radially symmetric field of a Pearl vortex. 
Conversely, if the skyrmion is displaced to a point $\bm{a}\neq 0$, we define this as an eccentric state.

The paper's structure is as follows. Section~\ref{Sec:GenAnsatz} formulates the method developed for studying a deformed skyrmion in an inhomogeneous magnetic field. In particular, Subsection~\ref{Sec:Model} introduces the model of a N\'eel-type skyrmion in an inhomogeneous magnetic field. 
Subsection~\ref{sec:ansatz} formulates the generalized \textit{Ansatz} for skyrmion deformation in any weak external magnetic field. 
Subsection~\ref{Sec:Coaxial} explains the theoretical approach for coaxial configurations, including a comprehensive description of the coaxial skyrmion \textit{Ansatz}, as informed by Ref.~\cite{Apostoloff2023}, and being the framework for the eccentric skyrmion \textit{Ansatz}. 
Subsection~\ref{Sec:Eccentric} is dedicated to detailed constructing the \textit{Ansatz} and calculation of the free energy to the second order approximation in weakness of the external field.
Subsection~\ref{Sec:OptimalConfig} outlines the method for determining optimal skyrmion parameters, such as its radius and effective domain wall width, and its optimal position relative to the non-uniform external field. 

The developments of Section~\ref{Sec:GenAnsatz} are then applied to a skyrmion in the stray field of a Pearl vortex in Section~\ref{Sec:Sk_V}. In this section, we delineate the structure of the field, perform the necessary calculations, and present the results, namely, the phase diagram for a skyrmion-vortex pair and the variation of the skyrmion parameters as a function of the vortex effective strength. Comparative analysis of the analytical results with micromagnetic simulations is presented in subsection~\ref{Sec:MMM}. Finally, the paper concludes with a summary of the findings in Section~\ref{Sec:Disc}.

\section{Magnetization \textit{Ansatz} for a deformed skyrmion in a weak external magnetic field \label{Sec:GenAnsatz}}

\subsection{Model: Skyrmion in ferromagnetic film\label{Sec:Model}}

We consider a thin chiral ferromagnet film, whose magnetic free energy is given by~\cite{Bogdanov1989}
\begin{eqnarray}
\mathcal{F}_{\text{magn}}[\bm{m}] &=&d_F \int d^2 \bm{r} \bigg\{ A (\nabla \bm{m})^2 + K(1- m_z^2) 
\notag \\
&&{}+ D \bigl [m_z \nabla \cdot \bm{m} - (\bm{m}\cdot \nabla) m_z \bigr ] \bigg\}.
\label{eq:MagFe}
\end{eqnarray}
Here, $\bm{m}(\bm{r})$ denotes the unit vector of the magnetization direction, $d_F$ is the film thickness, $A>0$ and $K>0$ represent the exchange and effective\footnote{The demagnetizing field contribution is included in the effective perpendicular anisotropy constant, $K{=}K_0{-}2\pi M_s^2$ \protect\cite{Menezes2019,Andriyakhina2021,Kuznetsov2022}.} perpendicular anisotropy constants, respectively, and $D>0$ is the Dzyaloshinskii--Moriya interaction constant (DMI). The magnetic free energy is normalized such that $\mathcal{F}_{\text{magn}}$ is zero for the ferromagnetic state with $m_z=1$.

In the absence of an external magnetic field, the free energy, Eq.~\eqref{eq:MagFe}, suggests the existence of skyrmions. A single free N\'eel-type skyrmion located at the coordinate origin can be described by the following magnetization profile,
\begin{equation}
\label{eq:m_free}
\bm{m}_{\rm Sk} = 
\bm{e}_{\bm r} \sin \theta (r)  + \bm{e}_z \cos \theta (r).
\end{equation}
Here, $\bm{e}_{r}$ and $\bm{e}_{z}$ are the unit vectors in the radial direction and along the $z$ axis (perpendicular to the interface), respectively. The skyrmion angle $\theta(r)$ satisfies the Euler-Lagrange equation (ELE), which is derived by minimizing the free energy~$\mathcal{F}_{\text{magn}}$ (see Appendix~\ref{app:ELE}),
\begin{eqnarray}
\label{eq:ELE_theta_0}
\dfrac{\ell_{w}^2}{r}\partial_r(r \partial_r\theta)-\frac{\ell_{w}^2+r^2}{2 r^2}\sin2 \theta    
+ 2\epsilon \frac{ \sin^2 \theta}{r/\ell_{w}} = 0,
\end{eqnarray}
where we introduce the dimensionless DMI parameter~$\epsilon$ and the domain wall width~$\ell_{w}$,
\begin{equation}
    \epsilon=D/2\sqrt{AK},
    \qquad
    \ell_{w} = \sqrt{A/K}.
    \label{eq:epsilon-def}
\end{equation}
The latter serves as a natural length scale in the problem.

Equation~\eqref{eq:ELE_theta_0} describes a N\'eel-type skyrmion under the following boundary conditions,
\begin{equation}
\theta(r\to\infty)=0,
\quad
\theta(r=0)=\chi\pi.
\label{eq:bound_theta0}
\end{equation} 
The first condition ensures that the magnetization at a distance from the skyrmion center is uniform. The second condition specifies that the magnetization at the skyrmion center is inverted relative to the uniform magnetization. The term $\chi=\pm1$ represents the skyrmion chirality.
It is important to note that a free skyrmion in an isolated ferromagnetic film exhibits only one chirality, determined by the sign of the DMI parameter, $\chi=\sgn(\epsilon)$. This paper considers only a positive DMI, $\epsilon>0$. Consequently, the only possible solution of Eq.~\eqref{eq:ELE_theta_0} for the free skyrmion of chirality $\chi=+1$ is stable. However, the skyrmion in the external inhomogeneous magnetic field can posses both chiralities, see details in Ref.~\cite{Apostoloff2023}.

The direct numerical solution of Eqs.~\eqref{eq:ELE_theta_0} and~\eqref{eq:bound_theta0} allows one to determine the skyrmion profile. It is well established that the direct solution is well approximated by the so-called 360$^{\circ}$ domain-wall \textit{Ansatz}, ${\theta(r)\approx\theta_{R\delta}(r)}$, 
\begin{equation}
\label{eq:ansatz_free}
\theta_{R\delta}(r)\equiv2\arctan\frac{\sinh(R/\delta)}{\sinh(r/\delta)}.
\end{equation}
Here, the parameter $R$ encodes two quantities: the skyrmion radius~$|R|$ and chirality~$\chi=\sgn(R)$, while $\delta$~represents the effective domain wall width. We refer to $R$ as the skyrmion radius, unless otherwise specified. The parameters $R$ and $\delta$ can be obtained by numerically minimizing the free energy $\mathcal{F}_{\text{magn}}$ using the domain-wall \textit{Ansatz}. 

The effect of the external inhomogeneous magnetic field~$\bm{B}(\bm{r})$ on the thin ferromagnetic film is captured by the Zeeman term, which is added to the free energy,
\begin{gather}
\mathcal{F}_{Z}[\bm{m},{\bm{B}}]  = - d_F \int d^2 \bm{r} M_s \bm{m}\cdot \bm{B}
|_{z=+0}
,
\label{eq:F-Sk-V-0}
\end{gather}
where $M_s$ denotes the saturation magnetization. In subsequent calculations, the external magnetic field is considered at the surface of the thin ferromagnetic film, i.e., at $z=+0$. Thus, we will omit the notation~$|_{z=+0}$ unless it is specifically required.

To determine the stable states of the ferromagnetic film in the presence of an external magnetic field $\bm{B}(\bm{r})$, one minimizes the total free energy,
\begin{equation}
\label{eq:F_tot}
\mathcal{F}_{\rm tot}[\bm{m},{\bm{B}}]=\mathcal{F}_{\text{magn}}[\bm{m}]+\mathcal{F}_{Z}[\bm{m},{\bm{B}}].
\end{equation}

The simplest stable state is the ``no-skyrmion'' configuration, where the ferromagnetic film remains skyrmion-free under the influence of the inhomogeneous magnetic field~$\bm{B}$. We consider the field to be weak, characterized by an effective strength~$\gamma$,
\begin{equation}
\label{eq:gamma-gen-def}
\gamma=M_s B_0/(2K)\ll1,
\end{equation}
where $B_0$ is the spatially constant characteristic magnitude\footnote{Note that there is an ambiguity in the definition of~$B_0$. It should be chosen to provide: (i) the reasonable estimation of the effective strength~$\gamma$ in Eq.~\eqref{eq:gamma-gen-def}, and (ii) vector~$\mu_{\bm{b}}$ in Eq.~\eqref{eq:mu_b} to be of the order of $\gamma^{0}$.} of the field~${\bm{B}=B_0\bm{b}}$ within the film.
The magnetization~$\bm{m}_{\bm{b}}$ of this stable state is then approximately given by
\begin{equation}
\label{eq:mu_b}
\bm{m}_{\bm{b}}\approx\bm{e}_z+\gamma\bm{\mu}_{\bm{b}},
\end{equation}
where $\bm{\mu}_{\bm{b}}(\bm{r})$ is a vector of the order of $\gamma^0$, orthogonal to~$\bm{e}_z$, i.e., $\bm{\mu}_{\bm{b}}\cdot \bm{e}_z=0$.

By expanding the total free energy in Eq.~\eqref{eq:F_tot} with~${\bm{m}=\bm{m}_{\bm{b}}}$ to second order in~$\gamma$ and minimizing it, one obtains the ELE for~$\bm{\mu}_{\bm{b}}$,
\begin{equation}
\label{eq:mu_b_ELE}
    \ell_w^2\Delta \bm{\mu}_{\bm{b}}-\bm{\mu}_{\bm{b}}+\bm{b}_{||}=0,
\end{equation}
where $\Delta$ denotes the Laplacian, and $\bm{b}_{||}(\bm{r})$ represents the in-plane component of the normalized external magnetic field~$\bm{b}(\bm{r})$,
\begin{equation}
    \bm{b}_{||}=\bm{b}-\bm{e}_z b_z.
\end{equation}

If the external magnetic field exhibits central symmetry, 
\begin{equation}
\label{eq:b-symmetr}
\bm{b}(\bm{r})={b}_{r}(r)\bm{e}_r+{b}_{z}(r)\bm{e}_z,
\end{equation}
then the vector~$\bm{\mu}_{\bm{b}}$ aligns with~$\bm{e}_{r}$,
\begin{eqnarray}
\label{eq:mu_theta}
    \bm{\mu}_{\bm{b}}(\bm{r})=\theta_{\bm{b}}(r)\bm{e}_{r},
\end{eqnarray}
and, given $\bm{b}_{||}={b}_{r}(r)\bm{e}_{r}$, Eq.~\eqref{eq:mu_b_ELE} reduces to
\begin{equation} \label{eq:thetagamma-ODE}
\dfrac{\ell_{w}^2}{r}\partial_r(r \partial_r{\theta}_{\bm{b}})-\frac{\ell_{w}^2+r^2}{r^2} {\theta}_{\bm{b}}  + {b}_{r} = 0.
\end{equation}
For a ``no-skyrmion'' configuration, with the radial component of the external field diminishing at infinity, the preceding equation is complemented by the following boundary conditions,
\begin{equation}
\theta_{\bm{b}}(r=0)=0,
\quad
\theta_{\bm{b}}(r\to\infty)=0.
\label{eq:bound_theta_b}
\end{equation} 

More complex stable states may involve the ferromagnetic film hosting one or several skyrmions. This paper focuses on configurations with a single skyrmion centered at a specific point $\bm a$. The analytic investigation of such states poses a significant challenge in determining the suitable magnetization profile~$\bm{m}$. One approach involves deriving the Euler-Lagrange equations (ELEs) and establishing boundary conditions where $\bm{m}=-\bm{e}_z$ at the skyrmion's center and $\bm{m}=\bm{e}_z$ at a distance. However, solving these ELEs, which are partial differential vector equations, demands computational resources comparable to those required for micromagnetic simulations. 

An alternative is to develop a simpler analytical \textit{Ansatz} that closely approximates the exact ELE solutions or micromagnetic simulation results. We introduce and detail such an \textit{Ansatz} for skyrmion magnetization under the influence of an external weak inhomogeneous magnetic field below.

\subsection{Formulation of the magnetization \textit{Ansatz} \label{sec:ansatz}}

In this subsection, we briefly outline the basic framework for constructing the \textit{Ansatz}. The comprehensive justification with some technical details for this approach will be provided in subsequent subsections, namely, the coaxial skyrmion \textit{Ansatz} in Subsection~\ref{Sec:Coaxial} and the eccentric skyrmion \textit{Ansatz} in Subsection~\ref{Sec:Eccentric}. An alternative representation of the \textit{Ansatz} using local coordinates is detailed in Appendix~\ref{app:LocalCoord}. For ease of subsequent construction, we consider the skyrmion centered at a point~$\bm{a}$ and shift the coordinate origin to this skyrmion center. Consequently, the external magnetic field is redefined as~${\bm{B}^{\bm{a}}(\bm{r})=\bm{B}(\bm{r}_{\bm{a}})}$, with ${\bm{r}_{\bm{a}}=\bm{r}+\bm{a}}$.

The core concept of the \textit{Ansatz} is to approximate the skyrmion magnetization~$\bm{m}_{\rm Sk}$ with a centrally symmetric unit-vector function as the leading term,
\begin{equation}
    \bar{\bm{m}}_{\rm Sk}=\bm{e}_{r} \sin \bar{\theta} (r) + \bm{e}_z \cos \bar{\theta} (r),
 \label{eq:bar-m}
\end{equation}
where $\bar{\theta} (r)$ represents the skyrmion angle\footnote{ Hereafter we use the bar sign over the functions to indicate that they are centrally symmetric and related to the averaged magnetic field $\bar{\bm{B}}^{\bm{a}}(r)$, see Eq.~\eqref{eq:B-aver}.}. The deformation from this symmetric state is considered only as a first-order correction in the small effective strength~$\gamma$ of the external magnetic field,
\begin{eqnarray}
	&&\bm{m}_{\rm Sk}\approx\bar{\bm{m}}_{\rm Sk}+\gamma\big[\bar{\bm{m}}_{\rm Sk}\times  [(\bm{\mu}_{\bm{b}^{\bm{a}}}-\bm{\mu}_{\bar{\bm{b}}^{\bm{a}}}) \times\bar{\bm{m}}_{\rm Sk}]\big]
 \label{eq:ansatz_eccentric}
 \\
 &&\quad=\bar{\bm{m}}_{\rm Sk}\{1-\gamma[(\bm{\mu}_{\bm{b}^{\bm{a}}}-\bm{\mu}_{\bar{\bm{b}}^{\bm{a}}})\cdot\bar{\bm{m}}_{\rm Sk}]\}+\gamma(\bm{\mu}_{\bm{b}^{\bm{a}}}-\bm{\mu}_{\bar{\bm{b}}^{\bm{a}}}).
 \notag
\end{eqnarray}
In this expression, $\bm{\mu}_{\bm{b}^{\bm{a}}}$ and $\bm{\mu}_{\bar{\bm{b}}^{\bm{a}}}$ are determined analogously to $\bm{\mu}_{\bm{b}}$ in Eq.~\eqref{eq:mu_b_ELE}, 
\begin{equation}
\notag
    \ell_w^2\Delta \bm{\mu}_{\bm{b}^{\bm{a}}}-\bm{\mu}_{\bm{b}^{\bm{a}}}+\bm{b}^{\bm{a}}_{||}=0,
    \quad
    \ell_w^2\Delta \bm{\mu}_{\bar{\bm{b}}^{\bm{a}}}-\bm{\mu}_{\bar{\bm{b}}^{\bm{a}}}+\bar{\bm{b}}^{\bm{a}}_{||}=0,
\end{equation}
with 
$%\bm{b}=
\bm{b}^{\bm{a}}=\bm{B}^{\bm{a}}/B_0$ and 
$%\bm{b}=
\bar{\bm{b}}^{\bm{a}}=\bar{\bm{B}}^{\bm{a}}/B_0$, 
respectively, and boundary conditions of Eqs.~\eqref{eq:bound_theta_b}. 
The function~$\bar{\bm{B}}^{\bm{a}}(\bm{r})$ is defined as the magnetic field averaged over the polar angle~$\phi$ around the skyrmion center, considering radial, azimuthal, and out-of-plane components,
\begin{eqnarray}
\label{eq:B-aver}
    \bar{\bm{B}}^{\bm{a}}=\langle B^{\bm a}_{r}\rangle_\phi\bm{e}_{r}+
    \langle B^{\bm a}_{z}\rangle_\phi\bm{e}_{z},
    \quad
    \langle \ldots\rangle_\phi
    \equiv\intl_{-\pi}^{\pi}\dfrac{d\phi}{2\pi}\ldots.
\end{eqnarray}

It should be noted that the magnetization ${\bm{m}_{\bar{\bm{b}}^{\bm{a}}}\approx\bm{e}_z+\gamma\bm{\mu}_{\bar{\bm{b}}^{\bm{a}}}}$ is determined by the centrally symmetric field~$\bar{\bm{B}}^{\bm{a}}$, and thus it maintains central symmetry, with $\bm{\mu}_{\bar{\bm{b}}^{\bm{a}}}$ being proportional to~$\bm{e}_{r}$ as indicated in Eq.~\eqref{eq:mu_theta}, i.e. $\bm{\mu}_{\bar{\bm{b}}^{\bm{a}}}(\bm{r})=\theta_{\bar{\bm{b}}^{\bm{a}}}(r)\bm{e}_{r}$.

\begin{figure}[t]
\centerline{\includegraphics[width=0.45\textwidth]{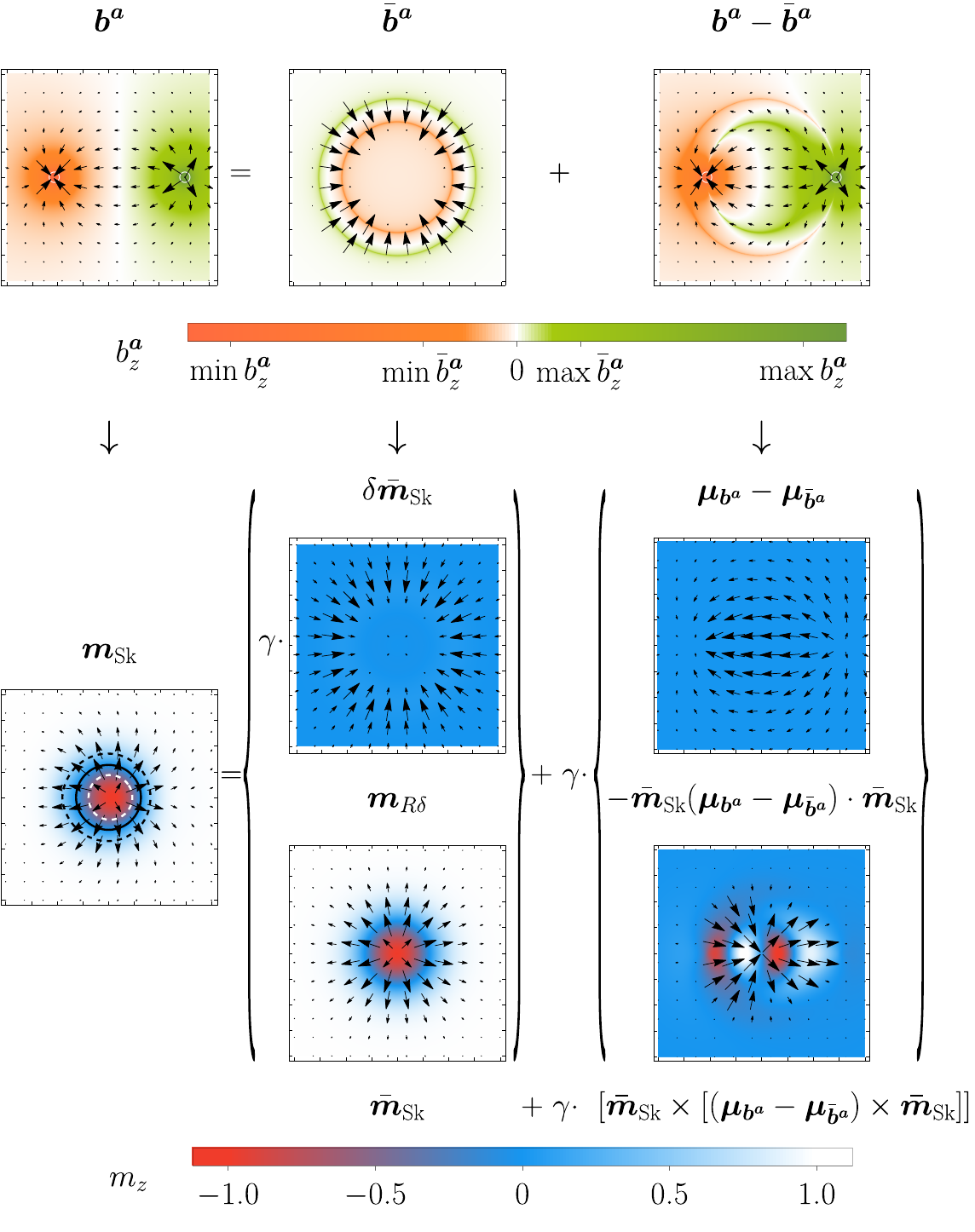}}
\caption{Visual representation of the \textit{Ansatz}~\eqref{eq:ansatz_eccentric}. For this illustrative example, we utilized the field generated by a pair consisting of a vortex (positioned on the left side of the topmost left plot, colored in orange) and an anti-vortex (located on the right side of the same plot, depicted in green tones). In each plot of the figure, distance coordinates are measured in units of the domain wall width, $\ell_w$, with the displayed region being $8 \ell_w \times 8 \ell_w$ in size. (\textit{First row}): decomposition of the total field $\bm{b}^{\bm{a}}$ acting on a skyrmion into (i) a cylindrically symmetric component $\bar{\bm{b}}^{\bm{a}}$, and (ii) the residual term $\bm{b}^{\bm{a}}-\bar{\bm{b}}^{\bm{a}}$. (\textit{Second row}): the total magnetization response $\bm{m}_{\rm Sk}$ (left column) is presented as a sum of several components. The middle column shows the cylindrically symmetric term, consisting of (\textit{bottom}) the domain wall \textit{Ansatz} $\bm{m}_{R\delta} \sim \gamma^0$, and (\textit{top}) the $\gamma^1$ correction $\gamma \delta \bar{\bm{m}}_{\rm Sk} = (\bar{\bm{m}}_{\rm Sk} - \bm{m}_{R\delta})$. Deviations from polar symmetry, also a $\gamma^1$ correction, are depicted in the right column and include (\textit{top}) the response of a ferromagnetic film (without a skyrmion) to the field $\bm{b}^{\bm{a}}-\bar{\bm{b}}^{\bm{a}}$, denoted by $\bm{\mu}_{\bm{b}^{\bm{a}}} - \bm{\mu}_{\bar{\bm{b}}^{\bm{a}}}$, and (\textit{bottom}) corrections due to the unit vector nature of total magnetization. To emphasize the breaking of cylindrical symmetry, contour plots are added to the leftmost figure of the second row, with the dashed black line for $m_z = 0.5$, the dashed white line for $m_z = -0.5$, and the solid black line for $m_z = 0$.}
\label{fig:ansatz}
\end{figure}

The skyrmion angle~$\bar\theta$ can be considered as an angle of a skyrmion in the centrally symmetric field~$\bar{\bm{B}}^{\bm{a}}$, and, consequently, it obeys the correspondent Euler-Lagrange equation. The exact solution of the mentioned equation can be approximated by the following coaxial \textit{Ansatz}, see Sec.~\ref{Sec:Coaxial} and~\cite{Apostoloff2023},
\begin{equation}
    \bar\theta(r)\approx\theta_{R\delta}^{\gamma \bm{a}}(r)\equiv
	\theta_{R\delta}(r)+\gamma\theta_{\bar{\bm{b}}^{\bm{a}}}(r)\cos\theta_{R\delta}(r),
 \label{eq:bar-ansatz}
\end{equation}
where $\theta_{R\delta}(r)$ represents the domain wall \textit{Ansatz} from Eq.~\eqref{eq:ansatz_free}, and $\theta_{\bar{\bm{b}}^{\bm{a}}}(r)$ is derived from Eq.~\eqref{eq:thetagamma-ODE} with $b_r=\bar{b}_r^{\bm{a}}$. The use of the \textit{Ansatz} for $\bar\theta(r)$, Eq.~\eqref{eq:bar-ansatz}, provides the way for the further analytic study. An explanatory visual representation of our proposed \textit{Ansatz} is illustrated in Figure~\ref{fig:ansatz}.
% and essentially reduces the computational costs. }

%%%%%%%%%%%%%%%%%%%%%%%%%%%%%%%%%%%
%FIGURE
%%%%%%%%%%%%%%%%%%%%%%%%%%%%%%%%%%%

Then one should calculate the total free energy given by Eq.~\eqref{eq:F_tot}, incorporating the \textit{Ansatz} according to Eqs.~\eqref{eq:ansatz_eccentric} and~\eqref{eq:bar-ansatz}, and minimize it with respect to unknown skyrmion parameters, %such as 
the radius~$R$ and effective wall width~$\delta$, and %to ascertain 
the stable skyrmion position $\bm{a}$. Typically, this involves 2D integration over the variable $\bm{r}$. However, Subsection~\ref{Sec:Eccentric} will present arguments supporting the \textit{Ansatz} in Eq.~\eqref{eq:ansatz_eccentric} and introduce simplified expressions for the total free energy that only require 1D radial integration, substantially lowering computational demands.

Further, we proceed to the detailed justifications of the \textit{Ansatz} introduced in that subsection and start our studying with the simplest case where the external magnetic field is centrally symmetric, and the skyrmion's center aligns with the field's center, as discussed in subsection~\ref{Sec:Coaxial}. In this scenario, the term~$\bar{\bm{m}}_{\rm Sk}$ alone persists in Eq.~\eqref{eq:ansatz_eccentric}. Building on the coaxial configuration findings, subsection~\ref{Sec:Eccentric} extends the discussion to the more complex scenario of an eccentric skyrmion.

\subsection{Coaxial skyrmion configuration\label{Sec:Coaxial}}

We commence the detailed analysis with the simpler case of a coaxial skyrmion in a centrally symmetric external magnetic field, as discussed in Ref.~\cite{Apostoloff2023}. This scenario corresponds to $\bm{a}=0$ and $\bar{\bm{B}}^0(\bm{r})=\bm{B}(\bm{r})$ as per the previous subsection. In such a configuration, the skyrmion is centrally symmetric, and the exact magnetization profile is sought as
\begin{equation}
\label{eq:m_coax}
\bm{m}_{\rm Sk} = 
\bm{e}_{\bm r} \sin \theta (r)  + \bm{e}_z \cos \theta (r).
\end{equation}

By minimizing the total free energy, Eq.~\eqref{eq:F_tot}, with the magnetization~$\bm{m}_{\rm Sk}$ from Eq.~\eqref{eq:bar-m}, we obtain the Euler-Lagrange equation for $\theta(r)$, detailed in Appendix~\ref{app:ELE},
\begin{eqnarray}
\label{eq:ELE_theta_coax}
&&\dfrac{\ell_{w}^2}{r}\partial_r(r \partial_r\theta)-\frac{\ell_{w}^2+r^2}{2 r^2}\sin2 \theta    
+ 2\epsilon \frac{ \sin^2 \theta}{r/\ell_{w}}
\notag \\
&& \qquad\qquad =\gamma  (b_z\sin\theta-b_{r}\cos\theta).
\end{eqnarray}

The boundary conditions for Eq.~\eqref{eq:ELE_theta_coax} remain the same, see Eq.~\eqref{eq:bound_theta0}, with identical implications: the chirality~$\chi$ takes the value of $\pm1$ for the skyrmion solution. It is important to note that, theoretically, both chiralities are possible for $\epsilon>0$ when influenced by an external magnetic field, as indicated by Ref.~\cite{Apostoloff2023}.

If the second boundary condition in Eqs.~\eqref{eq:bound_theta0} is modified to ${{\theta}(r=0)=0}$, the resulting solution of Eq.~\eqref{eq:ELE_theta_coax} characterizes the magnetization in a ``no-skyrmion'' state. For a weak magnetic field, characterized by $\gamma\ll1$, Eq.~\eqref{eq:ELE_theta_coax} can be linearized using $\theta(r)=\gamma \theta_{\bm{b}}$, leading to Eqs.~\eqref{eq:thetagamma-ODE} and~\eqref{eq:bound_theta_b}.

As demonstrated in Ref.~\cite{Apostoloff2023}, the exact solution of Eq.~\eqref{eq:ELE_theta_coax} under boundary conditions~\eqref{eq:bound_theta0} can be efficiently approximated with the following coaxial \textit{Ansatz},
\begin{equation}
\theta(r)\approx\theta_{R\delta}^{\gamma}(r)\equiv
\theta_{R\delta}(r)+\gamma{\theta}_{\bm{b}}(r)\cos\theta_{R\delta}(r).
\label{eq:coax-ansatz}
\end{equation}

The coaxial \textit{Ansatz} is predicated on the premise that the magnetization of an unperturbed skyrmion, represented by the angle~$\theta_{R\delta}$, rotates by an angle $\delta\theta$ due to the external magnetic field. At a distance from the skyrmion, where ${m_z\approx+1}$, this rotation angle is approximately $\delta\theta\approx+\gamma\theta_{\bm{b}}(r)$, akin to the ``no-skyrmion'' state. Conversely, near the skyrmion's center, where ${m_z\approx-1}$, the rotation is reversed, yielding $\delta\theta\approx-\gamma\theta_{\bm{b}}(r)$. Within the transitional region, $\delta\theta$ is proportionate to $\gamma\theta_{\bm{b}}(r)$ scaled by $m_z\approx\cos\theta_{R\delta}(r)$, leading to the formulation of the \textit{Ansatz} in Eq.~\eqref{eq:coax-ansatz}.

The \textit{Ansatz}~\eqref{eq:coax-ansatz} should be inserted into the free energy expression~\eqref{eq:F_tot}. Subsequently, the free energy must be minimized with respect to the two parameters $R$ and $\delta$. This method is faster and more efficient than directly solving Eq.~\eqref{eq:ELE_theta_coax} numerically. Furthermore, it yields relevant results across a broad range of the parameters $\epsilon$ and $\gamma$~\cite{Apostoloff2023}.

It is important to note that the rotation angle~$\delta\theta$ remains small when $\gamma\ll1$, making the \textit{Ansatz} in Eq.~\eqref{eq:coax-ansatz} seem as a minor variation of the free skyrmion. Nonetheless, this approach can account for significant changes in the skyrmion radius $R$ as well as alterations in chirality~\cite{Apostoloff2023}.

In this section, we have omitted the bar notation used in the previous subsection. However, the correspondence is straightforward for the reader to discern, considering that $b_{r/z}=\bar{b}^0_{r/z}$, $\theta(r)=\bar{\theta}(r)$, and ${\theta}_{\bm{b}}={\theta}_{\bar{\bm{b}}^0}$ within the coaxial setup.

\subsection{Total free energy for an eccentric skyrmion configuration\label{Sec:Eccentric}}

In this subsection, we expand the total free energy to first and second order in the small effective strength~$\gamma$ and minimize it analytically. Consequently, we derive formulas for the \textit{Ansatz} presented in subsection~\ref{sec:ansatz} and simplify the total free energy to the form of 1D integrals. These simplified expressions are intended for use in the numerical calculation of skyrmion parameters: radius~$R$, effective wall width~$\delta$, and position~$\bm{a}$. 

It is also instructive to outline the procedure for determining these parameters. As detailed in Appendix~\ref{app:F2-toy1}, to ascertain the parameters~$R$ and~$\delta$ with linear precision in $\gamma$, one must expand the free energy to first order in the small parameter $\gamma \ll 1$, as shown in Eq.~\eqref{eq:p0-p1}, and then minimize this expanded expression with respect to~$R$ and~$\delta$.  At this step, we can assume deformations of the centrally symmetric skyrmion profile to be negligible, but depending on~$\bm{a}$ via the averaged field~$\bar{\bm{b}}^a$. Formally speaking, we get the radius and effective wall thickness as a function of skyrmion position, $R(\bm{a})$ and $\delta(\bm{a})$.

Then, to accurately determine $\bm{a}$ with linear precision in $\gamma$, one must consider the quadratic terms in the free energy expansion, as indicated by Eq.~\eqref{eq:a1}. These terms involve the deformation part of the skyrmion magnetization. After substituting the functions $R(\bm{a})$ and $\delta(\bm{a})$ into this quadratic expansion, the total free energy is minimized with respect to $\bm{a}$.

\subsubsection{First order approximation}

For a weak inhomogeneous external magnetic field, characterized by $\gamma\ll1$, the leading approximation~$\bar{\bm{m}}_{\rm Sk}$ of the skyrmion magnetization~$\bm{m}_{\rm Sk}$ is a centrally symmetric unit vector function. The deformation due to the field manifests in the first-order approximation in $\gamma$. Consequently, we express the deformed magnetization as
\begin{equation}
\label{eq:ansatz_eccentric1}
    \bm{m}_{\rm Sk}\approx\bar{\bm{m}}_{\rm Sk}+\gamma\tilde{\bm{m}},
\end{equation}
where~$\bar{\bm{m}}_{\rm Sk}$ is defined by the skyrmion angle $\bar{\theta}$ in Eq.~\eqref{eq:bar-m}.

The deformation vector~$\tilde{\bm{m}}$ must be orthogonal to~$\bar{\bm{m}}_{\rm Sk}$ since both~${\bm{m}}_{\rm Sk}$ and~$\bar{\bm{m}}_{\rm Sk}$ are unit vectors. Therefore, $\tilde{\bm{m}}$ can be represented as a vector product,
\begin{equation}
	\label{eq:omega}
	\tilde{\bm{m}}=\bar{\bm{m}}_{\rm Sk}\times \bm{\omega},
\end{equation}
where $\bm{\omega}$ will be specified subsequently.

Note that any variation of~$\bar{\theta}$ on the order of the small parameter~$\gamma$ would significantly affect the first approximation~$\gamma\tilde{\bm{m}}$ in Eq.~\eqref{eq:ansatz_eccentric1}. Indeed, if it varies as $\bar\theta\to\bar\theta+\gamma\vartheta$, then the azimuthal component of $\bm{\omega}$ should be simultaneously modified, $\omega_\phi\to\omega_\phi+\vartheta$, in order to keep $\bm{m}_{\rm Sk}$ unchanged. To avoid this ambiguity, we stipulate that the skyrmion angle~$\bar{\theta}$ should be defined to the first order of small~$\gamma$ such that~$\omega_\phi$ (and, subsequently, the radial and out-of-plane components of~$\tilde{\bm{m}}$) is zero when averaged over the polar angle $\phi$,
\begin{equation}
\label{eq:aver_tilde_m}
\langle\omega_{\phi}\rangle_\phi=\langle\tilde{m}_{r}\rangle_\phi=\langle\tilde{m}_z\rangle_\phi=0.
\end{equation}

To determine the unknown skyrmion angle~$\bar{\theta}$ to the first order in $\gamma\ll1$, we insert the magnetization~${\bm{m}}_{\rm Sk}$ in the form of Eq.~\eqref{eq:ansatz_eccentric1} into the total free energy expression, Eq.~\eqref{eq:F_tot}, and expand it to the first order in~$\gamma$. This expansion reveals that the magnetization~${\bm{m}}_{\rm Sk}$ should be approximated by the leading term~$\bar{\bm{m}}_{\rm Sk}$,
\begin{equation}
	\label{eq:tot_expand1}
	\mathcal{F}_{\text{tot}}[{\bm{m}}_{\rm Sk},\bm{B}^{\bm{a}}]\approx
	\mathcal{F}_{\text{tot}}[\bar{\bm{m}}_{\rm Sk},\bm{B}^{\bm{a}}].
\end{equation}

In fact, Eq.~\eqref{eq:tot_expand1} should also include the first variation of the magnetic energy, 

\begin{eqnarray}
	&&\mathcal{F}_{\text{magn}}^{(1)}[{\bm{m}}_{\rm Sk},\gamma\tilde{\bm{m}}]
	=-2\gamma d_F \int d^2 \bm{r} \big[ A \tilde{\bm{m}}\cdot\Delta{\bm{m}}_{\rm Sk}  
		 \\\notag
		&&\quad{}+ K\tilde{m}_z\bar{m}_{{\rm Sk},z}- D \big(\tilde{m}_z \nabla \cdot {\bm{m}}_{\rm Sk} - \tilde{\bm{m}}\cdot \nabla \bar{m}_{{\rm Sk},z} \big) \big] ,
\end{eqnarray}
where $\Delta$ denotes the Laplacian. Integration over the polar angle~$\phi$ is necessary only for the radial and out-of-plane components of~$\tilde{\bm{m}}$, as ${\bm{m}}_{\rm Sk}$ is a centrally symmetric function. By considering Eqs.~\eqref{eq:aver_tilde_m}, one can verify that the first variation equals zero, $\mathcal{F}_{\text{magn}}^{(1)}[\gamma\tilde{\bm{m}},{\bm{m}}_{\rm Sk}]=0$, and thus it does not contribute to Eq.~\eqref{eq:tot_expand1}.

Upon integrating over the polar angle, it becomes apparent that the external field~$\bm{B}^{\bm{a}}$ can be replaced in Eq.~\eqref{eq:tot_expand1} by its azimuthally averaged counterpart~$\bar{\bm{B}}^{\bm{a}}$, as detailed in Eq.~\eqref{eq:B-aver}. Consequently, we arrive at an explicit formula for the total free energy,

\begin{eqnarray}
    &&
    \dfrac{\mathcal{F}_{\text{tot}}[{\bm{m}}_{\rm Sk},{\bm{B}}^{\bm{a}}]}{2\pi d_F A }
    \approx
    \dfrac{\mathcal{F}_{\text{tot}}[\bar{\bm{m}}_{\rm Sk},\bar{\bm{B}}^{\bm{a}}]}{2\pi d_F A }=
    \dfrac{\mathcal{F}_{\text{tot}}[\bm{m}_{\bm{b}},{\bm{B}}]}{2\pi d_F A }
\notag
\\ \notag
    &&\quad
    {}+ \int\limits_0^\infty \dfrac{dr \, r}{\ell_{w}^{2}} \Big\{ 
    \Big(\frac{\ell_{w}^2}{r^2}+  1\Big)\sin^2\bar\theta
    +2\epsilon \Big (\ell_{w}\partial_r\bar{\theta} +\frac{\sin2\bar\theta}{2r/\ell_{w}}\Big)
\\\label{eq:tot_expand1_explicit}
    &&\qquad
    {}+  \ell_{w}^{2}(\partial_r\bar{\theta})^2
    -2\gamma 
	\big[\bar{b}_{r}^{\bm{a}}\sin\bar\theta
        -2\bar{b}_z^{\bm{a}}
	\sin^2(\bar\theta/2)
	\big]
	\Big\} .
\end{eqnarray}

Here, the first term represents the free energy of the ``no-skyrmion" configuration, which is independent of any skyrmion parameters such as radius~$R$, effective wall width~$\delta$, or position~$\bm{a}$. Therefore, it can be disregarded during the minimization process.

\subsubsection{Equation and \textit{Ansatz} for $\bar{\theta}$}

Minimizing the total free energy, given by Eq.~\eqref{eq:tot_expand1_explicit}, with respect to the skyrmion angle $\bar{\theta}(r)$, we obtain ELE,
%the Euler-Lagrange equation,
\begin{eqnarray}
	\label{eq:ELE_theta_bar}
	&&\dfrac{\ell_{w}^2}{r}\partial_r(r \partial_r\bar\theta)-\frac{\ell_{w}^2+r^2}{2 r^2}\sin2 \bar\theta    
	+ 2\epsilon \frac{ \sin^2 \bar\theta}{r/\ell_{w}}
	\notag \\
	&& \qquad\qquad =\gamma  (\bar{b}_{z}^{\bm{a}}\sin\bar\theta-\bar{b}_{r}^{\bm{a}}\cos\bar\theta),
\end{eqnarray}
which is analogous to Eq.~\eqref{eq:ELE_theta_coax} for the coaxial case, but with $\bar{b}_{r/z}^{\bm{a}}$ replacing~$b_{r/z}$. 

Appropriate boundary conditions must accompany this equation, 
analogously to Eqs.~\eqref{eq:bound_theta0} and the related discussion,
\begin{equation}
	\bar\theta(r\to\infty)=0,
	\quad
	\bar\theta(r=0)=\chi\pi.
	\label{eq:bound_theta_bar}
\end{equation}

The approximate solution of Eq.~\eqref{eq:ELE_theta_bar} can be obtained similarly to that of Eq.~\eqref{eq:ELE_theta_coax} for the coaxial configuration. In fact, we should simply change $\bm{b}$ to $\bar{\bm{b}}^{\bm{a}}$ in Eq.~\eqref{eq:coax-ansatz} and get Eq.~\eqref{eq:bar-ansatz}.

Then, we employ the \textit{Ansatz} as specified in Eq.~\eqref{eq:bar-ansatz} and perform numerical minimization of the total free energy in Eq.~\eqref{eq:tot_expand1_explicit}. 
This process yields the optimal skyrmion parameters, radius~$R$ and effective wall width~$\delta$, for a given skyrmion position~$\bm{a}$. To ascertain the stable position~$\bm{a}$, we must evaluate the second-order approximation of the total free energy, as elucidated in Appendix~\ref{app:F2-toy1}.

\subsubsection{Second order approximation}

To determine the dependence of the skyrmion position~$\bm{a}$ on the effective strength~$\gamma$ of the magnetic field, it is necessary to minimize the total energy $\mathcal{F}_{\rm tot}[\bm{m}_{\rm Sk},\bm{B}^{\bm{a}}]$ with respect to~$\bm{a}$ as well. Since the position~$\bm{a}$, after redefining the coordinate origin to the skyrmion center, influences the total energy starting from the first order in $\gamma$, we compute $\mathcal{F}_{\rm tot}[\bm{m}_{\rm Sk},\bm{B}^{\bm{a}}]$ to the second order in $\gamma$ to accurately determine the function $\bm{a}(\gamma)$, as detailed in Appendix~\ref{app:F2-toy1}.

To achieve that, we extend the expansion of~$\bm{m}_{\rm Sk}$ to include terms up to the second order in~$\gamma$. Introducing a second-order term $\gamma^2\bm{\lambda}$ into Eq.~\eqref{eq:ansatz_eccentric1} and ensuring the normalization of $\bm{m}_{\rm Sk}$ to unity, we express the expansion up to the second order as
\begin{equation}
	\label{eq:ansatz_eccentric2}
	\bm{m}_{\rm Sk}\approx\bar{\bm{m}}_{\rm Sk}+\gamma\tilde{\bm{m}}+\gamma^2\big[\bm{\lambda} - (\bm{\lambda}\cdot\bar{\bm{m}}_{\rm Sk})\bar{\bm{m}}_{\rm Sk}
	- \bar{\bm{m}}_{\rm Sk}\tilde{m}^2/2\big].
\end{equation}

Upon substituting this form of $\bm{m}_{\rm Sk}$ into Eq.~\eqref{eq:F_tot}, we obtain the following expression for the total energy,
\begin{equation}
	\label{eq:tot_expand2}
	\mathcal{F}_{\text{tot}}[\bm{m}_{\rm Sk},\bm{B}^{\bm{a}}]\approx
	\mathcal{F}_{\text{tot}}[\bar{\bm{m}}_{\rm Sk},\bar{\bm{B}}^{\bm{a}}]+\mathcal{F}^{(2)}[\bar{\bm{m}}_{\rm Sk},\tilde{\bm{m}},\bm{a}],
\end{equation}
where the first term is as defined in Eq.~\eqref{eq:tot_expand1_explicit}, and the second term is given by
\begin{eqnarray}
\label{eq:F2}
&&\mathcal{F}^{(2)}[\bar{\bm{m}}_{\rm Sk},\tilde{\bm{m}},\bm{a}]
\notag
\\&&\qquad{}
=\mathcal{F}_{\text{tot}}'[\gamma\tilde{\bm{m}},\bm{B}^{\bm{a}}]
-\mathcal{F}_{\text{magn}}^{(1)}[\bar{\bm{m}}_{\rm Sk},\gamma^2\bar{\bm{m}}_{\rm Sk} \tilde{m}^2/2].
\qquad
\end{eqnarray}
In this equation, the prime notation in the first summand indicates that the calculation of~$\mathcal{F}_{\text{magn}}$, as per Eq.~\eqref{eq:MagFe}, within the total free energy formula~\eqref{eq:F_tot} excludes the constant term proportional to~$K$.

Additional terms from Eq.~\eqref{eq:ansatz_eccentric2} that contribute to $\mathcal{F}^{(2)}$ are of the order of~$\gamma^3$ and are therefore omitted. For a detailed explanation, refer to Appendix~\ref{app:F2-calc}. The validity of this approximation is confirmed below for a specific term only,
\begin{gather}
\dfrac{\mathcal{F}_{\text{magn}}^{(1)}[\bar{\bm{m}}_{\rm Sk},\gamma^2\bm{\lambda}(1-\bm{\lambda}\cdot\bar{\bm{m}}_{\rm Sk})]}{2\pi d_F A }
=2\gamma^2
\int \dfrac{d r\,r}{\ell_{w}^2} [\bm{\lambda}\times\bar{\bm{m}}_{\rm Sk}]_\phi
\notag\\\quad{}
\times
\Big[
\dfrac{\ell_{w}^2}{r}\partial_r(r \partial_r\bar\theta)-\frac{\ell_{w}^2+r^2}{2 r^2}\sin2 \bar{\theta}  
	+ 2\epsilon \frac{ \sin^2 \bar{\theta}}{r/\ell_{w}}
\Big].
\label{eq:F2-spare1}
\end{gather}

While this term may initially appear to be of the order of $\gamma^2$, the expression within the square brackets is actually of the order of $\gamma$, as indicated by Eq.~\eqref{eq:ELE_theta_bar}. Consequently, the term is effectively of the order of $\gamma^3$ and can be disregarded.

It is important to note that $\bar{\theta}$, as it enters $\mathcal{F}_{\text{tot}}[\bar{\bm{m}}_{\rm Sk},\bar{\bm{B}}^{\bm{a}}]$ within Eq.~\eqref{eq:tot_expand2}, should be computed only up to the first order, as specified by Eqs.~\eqref{eq:ELE_theta_bar} and Eq.~\eqref{eq:bar-ansatz}. Any second-order correction to $\bar{\theta}$, of the magnitude of $\gamma^2$, would only alter $\bm{\lambda}$ in Eq.~\eqref{eq:ansatz_eccentric2}, and thus would contribute to the free energy at an order of $\gamma^3$. Such a contribution is considered negligible and can be omitted.

\subsubsection{\textit{Ansatz} for skyrmion deformations \label{sec:ecc-ansatz2}}

As we can conclude from Eq.~\eqref{eq:F2} the minimization of $\mathcal{F}_{\text{tot}}[\bm{m}_{\rm Sk},\bm{B}^{\bm{a}}]$ should yield the Euler-Lagrange equations for the unknown vector function $\tilde{\bm{m}}$ or, equivalently, $\bm{\omega}$, see Eq.~\eqref{eq:omega}. As mentioned at the beginning of subsection~\ref{Sec:Eccentric}, directly solving such vector equations numerically is as intensive as micromagnetic modeling in terms of computer time and resources. To simplify calculations, we propose an \textit{Ansatz} for $\tilde{\bm{m}}$ that closely approximates the exact magnetization. This \textit{Ansatz} will subsequently be validated through micromagnetic modeling, as detailed in subsection~\ref{Sec:MMM}.

By expanding the leading approximation $\bar{\bm{m}}_{\rm Sk}$, which incorporates $\bar{\theta}$ from Eq.~\eqref{eq:bar-ansatz}, for small $\gamma$, we obtain
\begin{equation}
\label{eq:for-omega1}
  \bar{\bm{m}}_{\rm Sk}\approx \bm{m}_{R\delta}+\gamma[\bm{m}_{R\delta}\times  \bm{\mu}_{\bar{\bm{b}}^{\bm{a}}} \times  \bm{m}_{R\delta}],
\end{equation}
where $\bm{m}_{R\delta}$ is defined similarly to Eq.~\eqref{eq:m_free}, but with $\theta$ replaced by $\theta_{R\delta}$.

On the other hand, the magnetization~$\bm{m}_{{\bm{b}}^{\bm{a}}}$ in the ``no-skyrmion'' configuration can be represented as
\begin{eqnarray}
\label{eq:for-omega2}
    \bm{m}_{{\bm{b}}^{\bm{a}}}\approx \bm{e}_z+\gamma[\bm{e}_z\times  \bm{\mu}_{{\bm{b}}^{\bm{a}}} \times  \bm{e}_z].
\end{eqnarray}
By comparing Eqs.~\eqref{eq:for-omega1} and~\eqref{eq:for-omega2}, and considering that magnetization~${\bm{m}}_{\rm Sk}$ should include the position~$\bm{a}$ exclusively in the term $\gamma\bm{\mu}_{\bm{b}^{\bm{a}}}$, we formulate it as
\begin{equation}
\label{eq:for-omega3}
  {\bm{m}}_{\rm Sk}\approx \bm{m}_{R\delta}+\gamma[\bm{m}_{R\delta}\times  \bm{\mu}_{{\bm{b}}^{\bm{a}}} \times  \bm{m}_{R\delta}].
\end{equation}
Revisiting Eq.~\eqref{eq:ansatz_eccentric1} with Eq.~\eqref{eq:omega}, we define $\bm{\omega}$ as
\begin{equation}
	\bm{\omega}={\bm{\mu}}^{\bm{a}} \times\bar{\bm{m}},
 \qquad
{\bm{\mu}}^{\bm{a}}=
 \bm{\mu}_{\bm{b}^{\bm{a}}}-\bm{\mu}_{\bar{\bm{b}}^{\bm{a}}},
	\label{eq:omega2}
\end{equation}
which is consistent with Eqs.~\eqref{eq:for-omega3} and~\eqref{eq:ansatz_eccentric}.

The condition $\langle\omega_{\phi}\rangle_\phi=0$, as stated in Eq.~\eqref{eq:aver_tilde_m}, translates in the context of Eq.~\eqref{eq:omega2} to $\langle\mu^{\bm{a}}_r\rangle_\phi=0$ or, more explicitly,
\begin{equation}
\langle\bm{\mu}_{\bm{b}^{\bm{a}}}\cdot\bm{e}_r\rangle_\phi=
\bm{\mu}_{\bar{\bm{b}}^{\bm{a}}}\cdot\bm{e}_r=\theta_{\bar{\bm{b}}^{\bm{a}}}(r),
\end{equation}
which can be confirmed by comparing Eq.~\eqref{eq:thetagamma-ODE} with ${\bm{b}=\bar{\bm{b}}^a}$ and Eq.~\eqref{eq:mu_b_ELE} with $\bm{b}=\bm{b}^a$ after multiplying by $\bm{e}_r$ and averaging over~$\phi$.

\subsubsection{Explicit expression for $\mathcal{F}^{(2)}$}

Now, utilizing Eq.~\eqref{eq:omega2} for $\bm{\omega}$, we can integrate over the polar angle~$\phi$ in Eq.~\eqref{eq:F2} and present the explicit form for $\mathcal{F}^{(2)}$, 
\begin{gather}
	\dfrac{\mathcal{F}^{(2)}[\bar{\bm{m}}_{\rm Sk},\tilde{\bm{m}},\bm{a}]}{2\pi d_F A }=
	\gamma^2
	\int \dfrac{d r\,r}{\ell_{w}^2}  \Big\{ 
        2\tilde{b}_r^{\bm{a}} \sin^2 \bar{\theta}   
	+\tilde{b}_z^{\bm{a}}   \sin2 \bar{\theta}   
	\notag\\{}+
	[\tilde{\mu}^{\bm{a}}_{r}   \sin^2 \bar{\theta}   -\tilde{\mu}^{\bm{a}}_{\phi}   ]
	\ell_{w}^{2}(\partial_r\bar{\theta})^2  
	-\dfrac{\tilde{\mu}^{\bm{a}}_{r}   }{2r^2}
        (\ell_{w}^2+r^2)
	\sin^2 2\bar{\theta}   
	\notag\\\quad
	{}-2\epsilon \Bigl[
	\tilde{\mu}^{\bm{a}}_{\phi}   
	\ell_{w}\partial_r\bar{\theta}  
	+\frac{\tilde{\mu}^{\bm{a}}_{r\times\phi}   }{2r/\ell_{w}}\sin2\bar{\theta}   
	+\frac{\tilde{\mu}^{\bm{a}}_{r}   }{4r/\ell_{w}}\sin4\bar{\theta}   
	\Bigr]
	\Big\},
	\label{eq:F2-final}
\end{gather}
where we define certain functions considering position~$\bm{a}$ as a parameter,
\begin{gather}
        \tilde{b}_r^{\bm{a}}(r)=
	\big\langle{\mu}^{\bm{a}}_r\cdot(b_{r}^{\bm{a}}+ 
        \ell_{w}^2\Delta{\mu}^{\bm{a}}_r/2)\big \rangle_\phi
        -(\ell_{w}/r)^2\tilde{\mu}^{\bm{a}}_{r\times\phi}
        \qquad
        \notag\\
        \qquad\qquad\qquad\qquad
        {}
        +[(\ell_{w}/r)^2-1][\tilde{\mu}^{\bm{a}}_{r}   +\tilde{\mu}^{\bm{a}}_{\phi}   ]/2
	,
	\label{eq:tilde-func}\\\notag
	\tilde{b}_z^{\bm{a}}(r)=
	\big\langle {\mu}^{\bm{a}}_r{b}_z^{\bm{a}}\big\rangle_\phi,
        \;
        \tilde{\mu}^{\bm{a}}_{r}(r)=
	\big\langle({\mu}^{\bm{a}}_r)^2\big\rangle_\phi,
	\;
	\tilde{\mu}^{\bm{a}}_{\phi}(r)=
	\big\langle({\mu}^{\bm{a}}_{\phi})^2\big\rangle_\phi,
	\\\notag
	\tilde{\mu}^{\bm{a}}_{r\times\phi}(r)=\tilde{\mu}^{\bm{a}}_{r}   +\tilde{\mu}^{\bm{a}}_{\phi}+\Big\langle
	{\mu}^{\bm{a}}_r\partial_\phi{\mu}^{\bm{a}}_\phi
	-{\mu}^{\bm{a}}_\phi\partial_\phi{\mu}^{\bm{a}}_r
	\Big\rangle_\phi.
\end{gather}

\subsection{Optimal skyrmion configuration\label{Sec:OptimalConfig}}

Finally, we outline the procedure for determining the optimal skyrmion parameters -- radius~$R$, effective domain wall width~$\delta$, and position~$\bm{a}$ -- using the \textit{Ansatz} given by Eq.~\eqref{eq:ansatz_eccentric}. 

As briefly mentioned at the beginning of the previous section, the procedure involves taking the total free energy, as described by Eq.~\eqref{eq:tot_expand2}, and incorporating explicit expressions from Eqs.~\eqref{eq:tot_expand1_explicit} and~\eqref{eq:F2-final}. In these expressions, $\bar{\theta}$ is given as specified in Eq.~\eqref{eq:bar-ansatz}. Consequently, we express the total free energy as a function of $R$, $\delta$, and $\bm{a}$. Minimizing with respect to these parameters yields the stable skyrmion configuration, characterized by the radius $R$, effective domain wall width $\delta$, and position $\bm{a}$.

Furthermore, this procedure can be executed in two phases. Initially, the total free energy is minimized with respect to~$R$ and~$\delta$ alone. Subsequently, the resulting function of~$\bm{a}$ is analyzed to locate its minima, which correspond to the stable skyrmion positions. This approach differs from micromagnetic simulations in that it facilitates the identification of all potential metastable extremal states.

The numerical method described is notably efficient, utilizing solely 1D integration to compute the free energy. However, the accuracy of the results is confined to the regime where $\gamma\ll1$. A detailed comparison between the analytical findings and micromagnetic simulations will be presented in the following section.

\section{Skyrmion distortion due to a~Pearl vortex\label{Sec:Sk_V}}

In this part of the paper, we apply our generic findings to a specific problem: we study stable configurations of the N\'{e}el-type skyrmion in the stray field of a Pearl vortex.

\subsection{Stray field of a Pearl vortex}

We consider a heterostructure composed of superconducting and ferromagnetic films separated by a thin insulating layer, with the thickness being much smaller than the London penetration depth $\lambda_L$, effectively suppressing the proximity effect. The stray magnetic field generated by a Pearl vortex outside the superconducting film is given by \cite{Abrikosov-book}
\begin{gather}
{\bm B}_{V}
= \phi_0 \sgn(z) \nabla 
\int \frac{d^2\bm{q}}{(2\pi)^2} \frac{e^{-q |z| +i \bm{q}\cdot\bm{r} }}{q(1+2q\lambda)}.
\label{eq:vortex:B}
\end{gather}
Here, $\phi_0=h c/(2e)$ denotes the magnetic flux quantum, $\lambda=\lambda_L^2/d_S$ is the Pearl length~\cite{Pearl1964}, and the superconducting film thickness $d_S$ is assumed to be much smaller than the London penetration depth, $d_S\ll\lambda_L$. A more general formula for the magnetic field for arbitrary superconducting film thicknesses is available in Ref.~\cite{Carneiro2000} and detailed in Appendix~\ref{app:vortex}.

The magnetic field of the Pearl vortex exhibits central symmetry around the vortex core and is represented in the ferromagnetic film (at $z=+0$) as follows,
\begin{eqnarray}
{\bm B}_V\big|_{z=+0}=  
\frac{\phi_0}{4\pi \ell_{w}\lambda}
\left[
b_r(r)\bm{e}_{r}+b_z(r)\bm{e}_z
\right], \label{eq:vortex:B:thin-F}
\end{eqnarray}
where the functions $b_z(r)$ and $b_r(r)$ are specified by Eq.~\eqref{eq:vortex:h} in Appendix~\ref{app:vortex}.
 
Note that the characteristic magnitude of the Pearl vortex stray field is $B_0=\phi_0/(4\pi \ell_{w}\lambda)$, and thus the vortex effective strength~$\gamma$, as defined in Eq.~\eqref{eq:gamma-gen-def}, is given by
\begin{equation}
\label{eq:gamma-def}
\gamma=\dfrac{M_s\phi_0}{8\pi\lambda \sqrt{AK}}.
\end{equation}
In subsequent analytical studies, we will consider $\gamma$ to be small, which is consistent with experimental observations~\cite{Apostoloff2023}.

In this study, we examine the following range of model parameters,
\begin{equation}
d_F\sim d_S\ll \lambda_L \ll \ell_{w} \sim\delta\sim |R| \ll \lambda.
\label{eq:assumptions}
\end{equation} 
Within these assumptions, the functions $b_{r/z}(r)$ can be approximated as described in Appendix~\ref{app:vortex} and detailed in Ref.~\cite{Apostoloff2023}\footnote{The expressions for $b_{r/z}(r)$ provided herein are inverted in sign compared to those in Ref.~\cite{Apostoloff2023} due to differing normalization choices.}.

\begin{eqnarray}
b_z(r)= b_r(r)= -\ell_{w}/r , \qquad 
\lambda_L \ll r \ll \lambda .
\label{eq:vortex:B:b0_b1}
\end{eqnarray}
We note that the behavior of the vortex magnetic field at both short distances, $r\ll\lambda_L$, and large distances, $r\gg \lambda$, is not important for the physics within the parameter range of conditions~\eqref{eq:assumptions}.

\subsection{Skyrmion configurations}

Now we can apply the methodology developed in Section~\ref{Sec:GenAnsatz} to a N\'eel skyrmion under the influence of the stray field from a Pearl vortex. 

The coaxial configurations for skyrmion with both chiralities, $\chi=\pm1$,  are detailed in Ref.~\cite{Apostoloff2023}. This study concentrates on eccentric configurations with positive skyrmion chirality, $\chi=+1$, as such configurations are feasible only for this chirality sign, while stable configurations with $\chi=-1$ are coaxial only. Qualitatively this phenomenon was predicted in Ref.~\cite{Andriyakhina2021} for the nearly free skyrmion, here we only mention the reason. The free energy of the skyrmion with chirality~$\chi$ can be presented in the form $\mathcal{F}=\mathcal{F}_+-\chi\mathcal{F}_-$, where $\mathcal{F}_\pm$ are monotonic functions of the distance between centers of skyrmion and vortex. When $\chi=-1$, the function $\mathcal{F}(a)$ is monotonic, and the only minimum is possible in $a=0$. In the opposite case, $\chi=+1$ two monotonic functions are subtracted, which can produce a function with several minima.

To minimize the total free energy as described by Eq.~\eqref{eq:tot_expand2}, it is necessary to compute the functions related to the external magnetic field~$\bm{B}_V$ preliminarily. With the assumptions listed in Eq.~\eqref{eq:assumptions}, some of these functions are amenable to analytical calculation, as detailed in Appendix~\ref{app:averaged_vortex}, specifically,
\begin{eqnarray}
   \label{eq:thetagamma-approx}
    \bm{b}&=& -\dfrac{\bm{e}_{r}+\bm{e}_z}{r/\ell_{w}}, 
    \quad
    \theta_{\bm{b}}(r)=K_1(r/\ell_{w})-\ell_{w}/r,
\\
    \bm{b}^{\bm{a}}&=& -\dfrac{\bm{e}_{\bm{r}_{\bm{a}}}+\bm{e}_z}{r_{\bm{a}}/\ell_{w}},   
    \quad
    \bm{\mu}_{\bm{b}^{\bm{a}}}=\theta_{\bm{b}}(r_{\bm{a}})\bm{e}_{\bm{r}_{\bm{a}}},
 \\
    \bar{\bm{b}}^{\bm{a}}&=& \bar{b}_{r}^{\bm{a}}(r)\bm{e}_{r}+\bar{b}_{z}^{\bm{a}}(r)\bm{e}_z,  
    \quad
    \bm{\mu}_{\bar{\bm{b}}^{\bm{a}}}=\theta_{\bar{\bm{b}}^{\bm{a}}}(r)\bm{e}_{r},
\\
    \bar{b}_{r}^{\bm{a}}(r)&=& -\frac{\Theta(r-a)}{r/\ell_{w}},
    \quad
    \bar{b}_{z}^{\bm{a}}(r)=-\frac{K[4 a r/(a+r)^2] }{\pi  (a+r)/2\ell_{w}},
    \qquad
	\label{eq:vortex:field0}
 \\
 \label{eq:thetaAgamma-approx}
	\theta_{\bar{\bm{b}}^{\bm{a}}}(r)&=&\big[I_0(a/\ell_{w})K_1(r/\ell_{w})-\ell_{w}/r\big]\Theta(r-a)
	\notag\\&&{}
	-K_0(a/\ell_{w})I_1(r/\ell_{w})\Theta(a-r), 
\end{eqnarray}

Here, $\bm{e}_{\bm{r}_{\bm{a}}}=\bm{r}_{\bm{a}}/r_{\bm{a}}$ is a unit vector in the radial direction with respect to the shifted center, ${\bm{r}_{\bm{a}}=\bm{r}+\bm{a}}$, of the Pearl vortex (for the explanation of the shift, see the beginning of subsection~\ref{sec:ansatz}). The functions $I_{n}(z)$ and $K_n(z)$ are the modified Bessel functions of the first and second kinds, respectively, $K[z]$ is the complete elliptic integral of the first kind, and $\Theta(z)$ is the Heaviside step function.

The other functions, $\tilde{b}_{r/z}^{\bm{a}}$, $\tilde{\mu}^{\bm{a}}_{r/\phi}$, and $\tilde{\mu}^{\bm{a}}_{r\times\phi}$, must be calculated numerically from Eq.~\eqref{eq:tilde-func}. Then, the total free energy can be minimized using the form given in Eq.~\eqref{eq:tot_expand2} with explicit expressions from Eqs.~\eqref{eq:tot_expand1_explicit} and~\eqref{eq:F2-final}, applying $\bar{\theta}$ as per Eq.~\eqref{eq:bar-ansatz}. 

Due to the radial symmetry of the Pearl vortex, the skyrmion position~$\bm{a}$ enters the total free energy only as its magnitude~$a$, i.e., the distance between the centers of the skyrmion and the vortex. The minimization is thus by the three skyrmion parameters: radius~$R$, effective domain wall width~$\delta$, and distance~$a$ to the Pearl vortex center. The stable distance $a=0$ corresponds to the coaxial configuration, as studied in Ref.~\cite{Apostoloff2023}. The eccentric skyrmion configurations, where $a\neq0$, are detailed in the following subsection.

\subsection{Results of analytic approach}

In this section, we showcase the results derived from minimizing the total free energy, as outlined in Eq.~\eqref{eq:tot_expand2}, while employing the \textit{Ansatz} for $\bar{\theta}$ described in Eq.~\eqref{eq:bar-ansatz}.

\subsubsection{Skyrmion radius and distance in an eccentric configuration\label{sec:aRvsgamma}}

\begin{figure}[t]
\includegraphics[width=0.45\textwidth]{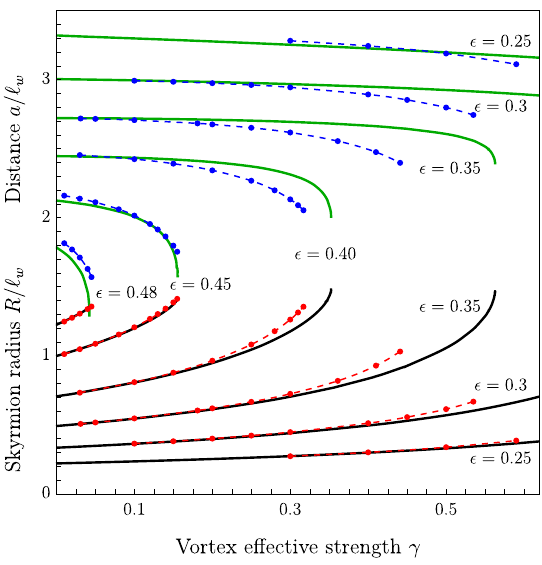}
\caption{The black and green solid lines present the dependence of the skyrmion radius $R/\ell_w$ and the distance $a/\ell_w$ between the centers of the skyrmion and vortex on the vortex effective strength $\gamma$, obtained by the minimization of the total free energy, Eq.~\eqref{eq:tot_expand2}, with the \textit{Ansatz} for $\bar{\theta}$, Eq.~\eqref{eq:bar-ansatz}. The numbers near the curves correspond to the values of the DMI parameter $\epsilon$. The circles mark the results extracted from the micromagnetic simulations.}
\label{fig:aRvsgamma}
\end{figure}

Figure~\ref{fig:aRvsgamma} presents the skyrmion radius~$R$ (lower group of black curves) and the distance~$a$ (upper group of green curves) between the centers of the skyrmion and the vortex in stable eccentric configurations as functions of the vortex effective strength~$\gamma$ for various values of the DMI parameter~$\epsilon$ (indicated by the numbers near the corresponding curves). The solid black and green curves result from minimizing Eq.~\eqref{eq:tot_expand2} using Eq.~\eqref{eq:bar-ansatz}, while the red and blue circles are derived from the outcomes of micromagnetic simulations (see Subsection~\ref{Sec:MMM}). 

%%%%%%%%%%%%%%%%%%%%%%%%%%%%%%%%%%%
%FIGURE
%%%%%%%%%%%%%%%%%%%%%%%%%%%%%%%%%%%

All curves initiate with a gentle slope at small~$\gamma$ and culminate in an abrupt increase for the radius~$R$ and an abrupt decrease for the distance~$a$. This behavior aligns with a square-root dependency near a certain critical value~$\gamma_{\rm cr}^+$, where $|\gamma_{\rm cr}^+-\gamma|\ll\gamma_{\rm cr}^+$. For further details, see the subsequent subsection~\ref{sec:Fa} and Appendix~\ref{app:F2-toy2},
\begin{eqnarray}
    R(\gamma_{\rm cr}^+)-R(\gamma)\propto a(\gamma)-a(\gamma_{\rm cr}^+)\propto \sqrt{\gamma_{\rm cr}^+-\gamma}.
\end{eqnarray}

An interesting feature can be observed in Fig.~\ref{fig:aRvsgamma}: the values of the skyrmion radius and the distance between the centers of the skyrmion and vortex draw close as $\gamma$ approaches $\gamma_{\rm cr}^+$. This peculiarity can be understood physically as follows. As the vortex effective strength~$\gamma$ increases, the skyrmion radius~$R$ expands due to the influence of the stray magnetic field. The larger the radius, the smaller the distance~$a$ must be, as can be deduced from the Zeeman energy, Eq.~\eqref{eq:F-Sk-V-0}, even for a free skyrmion. When~$R$ is significantly greater than~$a$, it implies that the skyrmion only contacts the vortex center with its tail, where the magnetization is approximately vertical, $\bm{m}=\bm{e}_z$. As the radius approaches the distance, the skyrmion's core, where the magnetization significantly deviates from the vertical, begins to overlap with the vortex center. The concurrent sharp changes in magnetization and the stray magnetic field in the same region are energetically unfavorable, leading the skyrmion to position itself directly above the vortex center. The subsequent subsection will delve into the dependence of the free energy on the distance~$a$ in greater detail.

\subsubsection{Dependence of the free energy on distance: several minima\label{sec:Fa}}

In this subsection, we examine the dependence of the free energy on the distance~$a$ to ascertain whether an eccentric or coaxial configuration is more favorable, given specific values of the parameters~$\epsilon$ and~$\gamma$. To this end, we introduce the function~$\mathfrak{F}(a)$ as
\begin{equation}
\label{eq:Fa-aver-Rdelta}
    \mathfrak{F}(a)\equiv \min_{R,\delta} \dfrac{\mathcal{F}_{\text{tot}}[\bm{m}_{\rm Sk},\bm{B}_V^{\bm{a}}]}{2\pi d_F A},
\end{equation}
which represents the total free energy $\mathcal{F}_{\text{tot}}[\bm{m}_{\rm Sk},\bm{B}_V^{\bm{a}}]$ from Eq.~\eqref{eq:tot_expand2} with $\bar{\theta}$ from Eq.~\eqref{eq:bar-ansatz}, normalized by~$2\pi d_F A$ and minimized with respect to the skyrmion size parameters, namely the radius~$R$ and effective domain wall width~$\delta$. A minimum of~$\mathfrak{F}(a)$ at a specific point $a=a_{\rm min}$ indicates a potential stable skyrmion configuration within the stray field of the Pearl vortex.

\begin{figure}[t]
\includegraphics[width=0.45\textwidth]{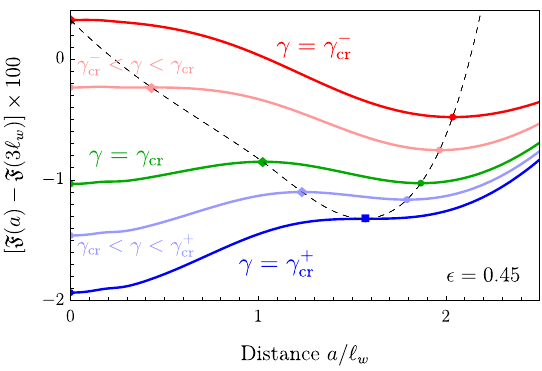}
\caption{The plot of the function $[\mathfrak{F}(a)-\mathfrak{F}(3\ell_w)]\times 100$, produced from the total free energy, Eq.~\eqref{eq:Fa-aver-Rdelta}, for the DMI parameter $\epsilon=0.45$ and several values of the vortex effective strength $\gamma$: $\gamma_{\rm cr}^{-}\approx0.084$; $0.115$; $\gamma_{\rm cr}\approx0.138$; $0.147$; $\gamma_{\rm cr}^{+}\approx0.156$ (from top to bottom). The circles, diamonds, and squares indicate the minima, maxima, and inflection points, respectively. The dashed curve continuously marks the positions of minima and maxima when varying $\gamma$. All plotted functions exhibit a monotonous increase in the hidden region, where $a > 3\ell_w$.}
\label{fig:F(a)}
\end{figure}

Figure~\ref{fig:F(a)} illustrates the variation of $\mathfrak{F}(a)$ for the DMI parameter $\epsilon=0.45$ across several values of $\gamma$. The subsequent discussion provides a more comprehensive categorization of potential types of dependence, based on the critical values $\gamma_{\rm cr}^{-}(\epsilon)$, $\gamma_{\rm cr}(\epsilon)$, and $\gamma_{\rm cr}^{+}(\epsilon)$.

%%%%%%%%%%%%%%%%%%%%%%%%%%%%%%%%%%%
%FIGURE
%%%%%%%%%%%%%%%%%%%%%%%%%%%%%%%%%%%

When $\epsilon<\epsilon_{\rm cr}^{-}\approx0.488$ and $\gamma<\gamma_{\rm cr}^{-}(\epsilon)$, the function~$\mathfrak{F}(a)$ exhibits a single minimum at $a=a_{\rm min}>0$, indicating that the only viable skyrmion--vortex configuration is eccentric.

When $\gamma>\gamma_{\rm cr}^{+}(\epsilon)$ or $\epsilon>\epsilon_{\rm cr}^{+}\approx0.493$, the function~$\mathfrak{F}(a)$ displays a solitary minimum at $a=0$, signifying that the coaxial skyrmion--vortex configuration is the sole possibility.

In the intermediate regime, where $\gamma_{\rm cr}^{-}(\epsilon)<\gamma<\gamma_{\rm cr}^{+}(\epsilon)$, the heterostructure can support both coaxial and eccentric skyrmion--vortex configurations, as the function~$\mathfrak{F}(a)$ presents at least two minima: one at $a=0$ and another at $a=a_{\rm min}>0$. Nonetheless, the free energy levels of these configurations differ, with the global minimum occurring for the coaxial configuration when $\gamma_{\rm cr}(\epsilon)<\gamma<\gamma_{\rm cr}^{+}(\epsilon)$, and for the eccentric configuration when $\gamma_{\rm cr}^-(\epsilon)<\gamma<\gamma_{\rm cr}(\epsilon)$.

It is noteworthy that the function~$\mathfrak{F}(a)$ may potentially exhibit more than two minima. Specifically, an additional local minimum at $a=a_{\rm add}\sim0.3\ell_w$ has been identified. However, for all values of $\epsilon$ and $\gamma$, this minimum is only local and its significance falls outside the accuracy of our second-order approximation for $\gamma\ll1$. Consequently, the analytic approach alone cannot guarantee the existence of such a stable configuration. Moreover, micromagnetic simulations have not confirmed the presence of this configuration, as detailed in subsection~\ref{Sec:MMM}.

The dashed line in Fig.~\ref{fig:F(a)} continuously traces the positions of the minima and maxima as $\gamma$ varies. It is evident that near $\gamma\approx\gamma_{\rm cr}^{+}$, the location of the minimum $a_{\rm min}$ changes rapidly with $\gamma$, converging with the maximum point $a_{\rm max}$ at the inflection point $a_{\rm in}$ precisely at $\gamma=\gamma_{\rm cr}^{+}$. This swift change leads to a sharp decrease (increase) in the distance $a$ (radius $R$), as shown in Fig.~\ref{fig:aRvsgamma} and explained in subsection~\ref{sec:aRvsgamma} and Appendix~\ref{app:F2-toy2}.

The distance $a$ is highly susceptible to the second-order terms detailed in Eq.~\eqref{eq:F2-final}, as elaborated in Appendix~\ref{app:F2}. This precision in calculation is reflected in the close correspondence between the curves for $a(\gamma)$ in Fig.~\ref{fig:aRvsgamma}. However, this susceptibility also implies that potential corrections to the free energy, arising from effects not accounted for and comparable in magnitude to the second-order approximation, could influence the experimental distance between the skyrmion and the vortex.

\subsubsection{Phase diagram}

\begin{figure}[t]
\includegraphics[width=0.45\textwidth]{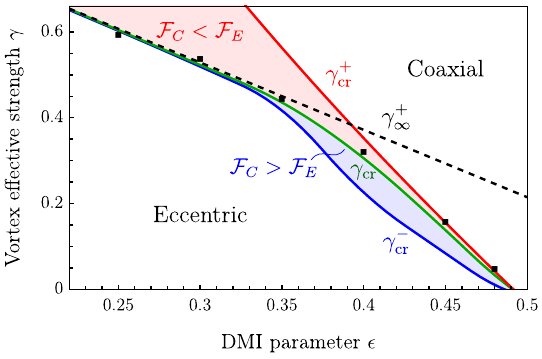}
\caption{Phase diagram, produced per the total free energy of Eq.~\eqref{eq:tot_expand2} with the \textit{Ansatz} for $\bar{\theta}$, Eq.~\eqref{eq:bar-ansatz}. The blue, green, and red solid curves show $\gamma_{\rm cr}^-(\epsilon)$, $\gamma_{\rm cr}(\epsilon)$, and $\gamma_{\rm cr}^+(\epsilon)$, respectively, delineating the separation of the phases from below to the top. The shaded areas represent the phases where the coaxial and eccentric states can coexist. The black dashed straight line denotes $\gamma_{\infty}^+=1-\pi\epsilon/2$, indicating the condition under which the coaxial skyrmion radius is comparable to the Pearl length. Black squares denote $\gamma_{\rm cr}(\epsilon)$ values obtained from micromagnetic simulations.}
\label{fig:phasediag}
\end{figure}

The findings from the previous subsection are represented in the phase diagram on the $(\epsilon,\gamma)$ plane, illustrated in Fig.~\ref{fig:phasediag}. This diagram delineates four distinct phases. The two unshaded regions indicate phases where exclusively eccentric configurations for $\gamma<\gamma_{\rm cr}^-(\epsilon)$, or solely coaxial configurations for $\gamma>\gamma_{\rm cr}^+(\epsilon)$, are viable. The two shaded regions denote phases where both eccentric and coaxial configurations are theoretically feasible within the same heterostructure, as the free energy exhibits two minima with respect to the distance $a$. In practice, both configurations have been observed in micromagnetic simulations around $\gamma\approx\gamma_{\rm cr}$, as detailed in subsection~\ref{Sec:MMM}.

%%%%%%%%%%%%%%%%%%%%%%%%%%%%%%%%%%%
%FIGURE
%%%%%%%%%%%%%%%%%%%%%%%%%%%%%%%%%%%

Note that the curves $\gamma_{\rm cr}^{-}(\epsilon)$ and $\gamma_{\rm cr}(\epsilon)$ approach the asymptotic line $\gamma_{\infty}^{+}(\epsilon)=1-\pi\epsilon/2$ for $\epsilon\lesssim0.3$. As the vortex effective strength $\gamma$ nears the critical threshold $\gamma_{\infty}^{+}$, detailed in Ref.~\cite{Apostoloff2023}, the radius $R$ of the coaxial skyrmion increases substantially, becoming comparable to the Pearl length $\lambda$, which is much larger than $\ell_{w}$. Consequently, the free energy of such an expanded coaxial skyrmion is markedly lower than that of any eccentric configuration. Given that $\gamma_{\rm cr}^{-}(\epsilon)$ and $\gamma_{\rm cr}^{+}(\epsilon)$ delineate the transition to coaxial configurations and the point where the energies of coaxial and eccentric configurations equalize, both critical values naturally fall below $\gamma_{\infty}^{+}$. The black squares in Fig.~\ref{fig:phasediag} represent the values of $\gamma_{\rm cr}(\epsilon)$ ascertained from micromagnetic simulations, corroborating the analytical predictions with a high degree of concordance.

\subsubsection{Comparison between different approximations of the skyrmion profile}

We complete the analytic study presented in this subsection by comparing several possible form that can be used as an \textit{Ansatz} for the skyrmion magnetization. In particular, we take the magnetization~$\bm{m}$ in one of the following forms and minimize the total free energy with respect to the skyrmion parameters, radius~$R$ and effective domain wall thickness~$\delta$, and the distance~$a$ between skyrmion and vortex centers. These forms are
\begin{enumerate}
    \item $\bm{m}=\bm{m}_{\rm Sk}$, where $\bm{m}_{\rm Sk}$ is the full anisotropic \textit{Ansatz} given by Eq.~\eqref{eq:ansatz_eccentric}; 
    \item $\bm{m}=\bar{\bm{m}}_{\rm Sk}$, where $\bar{\bm{m}}_{\rm Sk}$ is the centrally symmetric leading approximation given by Eq.~\eqref{eq:bar-m};
    \item $\bm{m}=\bm{m}_{R\delta}$, where $\bm{m}_{R\delta}$ is the centrally symmetric domain wall approximation given by Eq.~\eqref{eq:m_free} with $\theta=\theta_{R\delta}$, see Eq.~\eqref{eq:ansatz_free};
    \item $\bm{m}=\bm{m}_{0}$, where $\bm{m}_{0}=\bm{m}_{R_0\delta_0}$ is the same domain wall approximation as in the previous item, but with $R$ and $\delta$ taken for the free skyrmion, $R=R_0$ and $\delta=\delta_0$ at $\gamma\to0$, and unchanged when minimizing total free energy only with respect to distance $a$. 
\end{enumerate}

\begin{figure}[t]
\centerline{\includegraphics[width=0.45\textwidth]{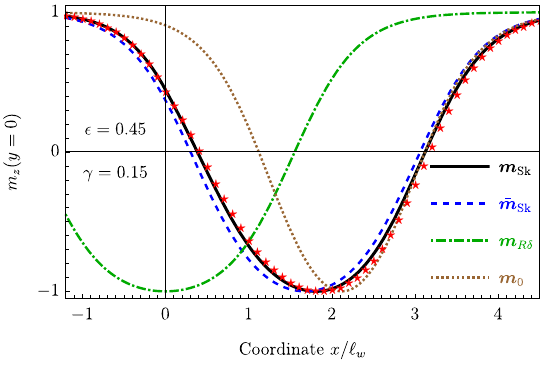}}
\caption{Distribution of $m_z(x,y=0)$ calculated through minimization of the total free energy with several forms of \textit{Ans\"atzes}, $\bm{m}_{\rm Sk}$ (solid black), $\bar{\bm{m}}_{\rm Sk}$ (dashed blue),  $\bm{m}_{R\delta}$ (dashed-dotted green), and $\bm{m}_{0}$ (dotted brown curves). The stars mark the values of $m_z(x,y=0)$ extracted from the micromagnetic simulation. }
\label{fig:comparison}
\end{figure}

Figure~\ref{fig:comparison} demonstrates the distribution $m_z(x,y=0)$ of skyrmion, when Pearl vortex is placed in coordinate origin. The solid black, dashed blue, dashed-dotted green, and dotted brown lines correspond to $\bm{m}_{\rm Sk}$,  $\bar{\bm{m}}_{\rm Sk}$, $\bm{m}_{R\delta}$, and $\bm{m}_{0}$ from the list above. The stars mark the values extracted from the micromagnetic simulation and they fit well by a solid black line of $\bm{m}_{\rm Sk}$.  
The values of $R$, $\delta$, and $a$, obtained from the minimization of the free energy, according to the list of magnetizations~$\bm{m}$ above, are presented in Table \ref{tab:ansatzes}
\begin{table}[!h]
    \centering
    \caption{The values of the radius $R$, the effective wall width $\delta$, and the distance $a$ corresponding to the several forms of \textit{Ans\"atze} shown in Fig.~\ref{fig:comparison}.}
    \begin{tabular}{|c|c|c|c|}
    \hline
    $\bm{m}$ & $R/\ell_w$  & $\delta/\ell_w$  & $a/\ell_w$ \\\hline
    $\bm{m}_{\rm Sk}$   & 1.37  & 0.841 & 1.76 \\\hline
    $\bar{\bm{m}}_{\rm Sk}$     & 1.38 & 0.842 & 1.69 \\\hline
    $\bm{m}_{R\delta}$     & 1.56 & 0.788 & 0 \\\hline
    $\bm{m}_{0}$    & 1.00 & 0.785 & 2.12 \\\hline
    \end{tabular}
    \label{tab:ansatzes}
\end{table}

%%%%%%%%%%%%%%%%%%%%%%%%%%%%%%%%%%%
%FIGURE
%%%%%%%%%%%%%%%%%%%%%%%%%%%%%%%%%%%

Note that solid black and dashed blue curves are close enough to each other, but the accounting for the non-symmetric terms in the full \textit{Ansatz}~$\bm{m}_{\rm Sk}$ (solid black) provides a more precise result than only the centrally symmetric leading approximation~$\bar{\bm{m}}_{\rm Sk}$ (dashed blue).
The coarse approximation with domain wall \textit{Ansatz} (dashed-dotted green) or even simply free skyrmion profile (dotted brown) yields inadequate results for the radius~$R$, thickness~$\delta$, and distance~$a$. In particular, for the sole domain wall \textit{Ansatz} and parameters used in Fig.~\ref{fig:comparison} the free energy has the only minimum in $a=0$. 

\subsection{Micromagnetic simulations \label{Sec:MMM}}

In support of our findings, we conducted a set of micromagnetic simulations using the Object Oriented MicroMagnetic Framework (OOMMF) \cite{OOMMF} and a Python library, Ubermag \cite{Ubermag}. We represent the system using classical magnetic moments positioned at the mesh centers, governed by magnetic interactions. To simulate an isolated magnet region, we applied periodic boundary conditions on the \(x\)-\(y\) plane.

To generate a skyrmion, we began with a region of reversed magnetization and allowed it to relax under the influences of Heisenberg exchange, Dzyaloshinskii-Moriya, and magnetic anisotropy interactions, and the vortex-induced magnetic field. In our simulations, we assumed a Pearl vortex with zero-sized core anchored at the grid's origin. As previously stated, we consider the vortex to be large, signified by $\lambda \gg \ell_{w}$, allowing its magnetic field to be approximated using Eqs.~\eqref{eq:vortex:B:thin-F}, with $b_z(r)$ and $b_r(r)$ detailed in Eqs.~\eqref{eq:vortex:B:b0_b1}. For other details of our micromagnetic analysis we refer to Appendix \ref{app:mmm}.

\begin{figure}[t]
\centerline{\includegraphics[width=0.45\textwidth]{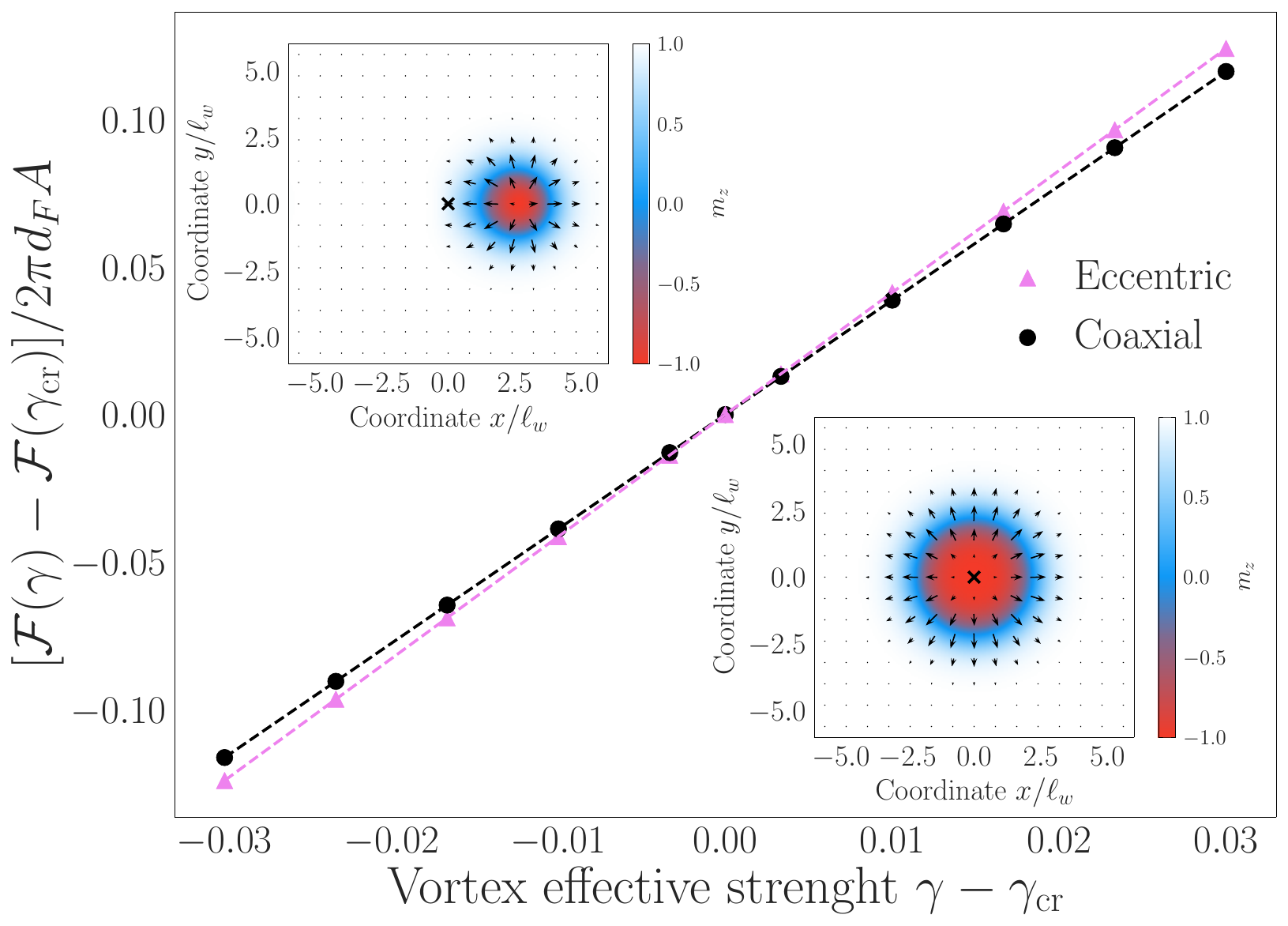}}
\caption{The energy dependencies of both the eccentric state (pink triangles) and the coaxial state (black circles) in proximity to the critical vortex force $\gamma_{\text{cr}}$ for $\epsilon = 0.4$. The plot and inset pictures are obtained from the micromagnetic simulations. In the vicinity of this critical point, the free energy exhibits nearly linear dependence on $\gamma - \gamma_{\rm cr}$. The intersection point of the two lines (pink and black) determines $\gamma_{\rm cr}(\epsilon)$, which is indicated by black dots in Figure~\ref{fig:phasediag}. We emphasize the remarkable agreement between the numerical data and the analytical predictions presented in the corresponding figure. The magnetization profiles of both states, eccentric and coaxial at $\gamma_{\rm cr}$, are provided in the figure's insets. The black cross located at the grid's center marks the position of the vortex.}
\label{fig:F0_Fa}
\end{figure}

This subsection addresses the configurations where the vortex and skyrmion are offset from one another. We explored the parameters $\epsilon$ and $\gamma$ to determine conditions that stabilize the shifted configuration. Fig. \ref{fig:phasediag} displays th possible stability regions for both eccentric and coaxial configurations. By initializing a skyrmion at the vortex's center and at an offset distance, we relaxed the system and compared the resulting energies. While displaced configurations corresponding to a local minimum were stabilized for $\gamma_{\rm cr} < \gamma < \gamma_{\rm cr}^+$, see Fig.~\ref{fig:F0_Fa}, we can confidently identify only the transition boundary at $\gamma_{\rm cr}$. The associated dots are illustrated using black squares in Fig.~\ref{fig:phasediag}. Consistent with theoretical predictions, our numerical experiments indicate that as $\epsilon$ increases, $\gamma_{\rm cr}$ decreases.

To quantify our simulations, using Eq.~\eqref{eq:ansatz_eccentric} for the skyrmion magnetization, we derived relationships between the parameters $R$ and $a$, and the strengths from the skyrmion ($\epsilon$) and vortex ($\gamma$). These results are summarized in Fig.~\ref{fig:aRvsgamma}. For weak vortex magnetic fields, skyrmion deformations were minor and its profile could be approximately represented by the domain wall \textit{Ansatz}, Eq.~\eqref{eq:ansatz_free}. However, with increasing $\gamma$, the deformation grew pronounced, rendering Eq.~\eqref{eq:ansatz_free} insufficient. In contrast, the effectiveness of our proposed \textit{Ansatz}, as defined in Eq.\eqref{eq:ansatz_eccentric}, relative to the domain wall \textit{Ansatz} and numerical simulations, is evident from Fig.\ref{fig:aRvsgamma} and Fig.~\ref{fig:phasediag}.
Indeed, the proposed \textit{Ansatz} accurately captures the evolution of the parameters $R$ and $a$ as functions of $\gamma$ and even demonstrates excellent performance in estimating $\gamma_{\rm cr}$, despite the seemingly large values.

%%%%%%%%%%%%%%%%%%%%%%%%%%%%%%%%%%%
%FIGURE
%%%%%%%%%%%%%%%%%%%%%%%%%%%%%%%%%%%

In addition, we describe a numerical experiment that demonstrates the possibility of stabilizing both coaxial and offset states with identical energy. As depicted in Fig.~\ref{fig:F0_Fa}, near $\gamma_{\rm cr}$, two different configurations with closely matched free energy values are observed, as shown in the insets depicting magnetization. In the first, the skyrmion is stabilized at a finite distance~$a$ (the upper inset), while in the second, it's centered (the lower inset). Intriguingly, the coaxial configuration is associated with a significant enlargement of the skyrmion 
radius -- nearly doubling. This observation aligns with the findings in Ref.~\cite{Apostoloff2023}, which elucidate the skyrmion radius expansion when a Pearl vortex is present.

%%%%%%%%%%%%%%%%%%%%%%%%%%%%%%%%%%%
%FIGURE
\begin{figure}[t]
\centerline{\includegraphics[width=0.45\textwidth]{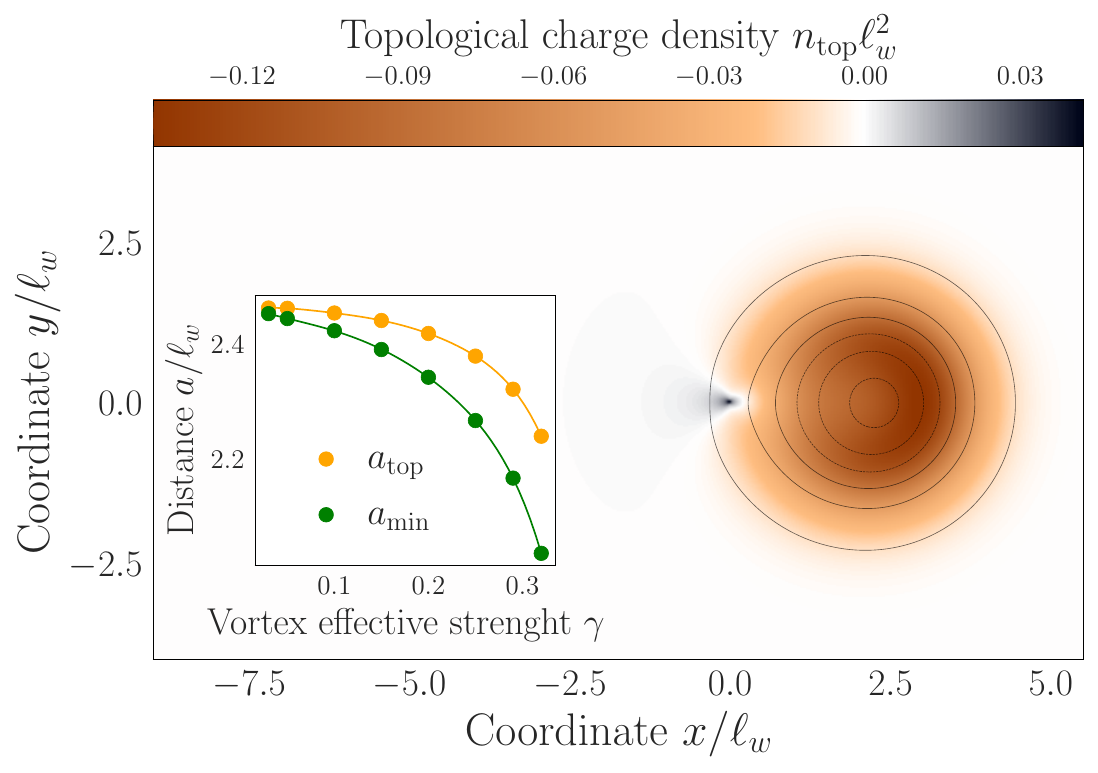}}
\caption{The topological charge density distribution for an eccentric skyrmion-vortex configuration obtained from the micromagnetic simulations. Material parameters are $\epsilon = 0.4$ and $\gamma = 0.3$. For comparison, we also include black contours representing the shape of the magnetization, ranging equidistantly from $m_z=-0.9$ (innermost) to $m_z=0.9$ (outermost). In the inset, we provide a calculation of the center of topological charge density using Eq.~\eqref{eq:a_top}. The plot also depicts the position of the magnetization center, $a_{\text{min}}$, as a function of $\gamma$. It is evident that $a_{\text{top}}$ and $a_{\text{min}}$ follow a similar trend as $\gamma$ increases, yet they differ from one another significantly.}
\label{fig:top_charge}
\end{figure}
%%%%%%%%%%%%%%%%%%%%%%%%%%%%%%%%%%%

Finally, we present a characteristic distribution of topological charge density, calculated using the formula
\begin{equation}\label{eq:top_density}
    n_{\text{top}}(\bm{r}) = \frac{1}{4\pi} \bm{m}(\bm{r}) \cdot \left( \frac{\partial \bm{m}(\bm{r})}{\partial x} \times \frac{\partial \bm{m}(\bm{r})}{\partial y} \right).
\end{equation}
While both the magnetization profile and the topological charge density of the skyrmion undergo deformations, notably, these deformations exhibit dissimilarity, as depicted in Fig. \ref{fig:top_charge}. In this instance, $a_{\text{top}}$ deviates from $a_{\text{min}}$, as illustrated in the inset of Fig. \ref{fig:top_charge}. Here, $a_{\text{top}}$ represents the center of the topological charge distribution, which can be determined using the formula
\begin{eqnarray}\label{eq:a_top}
    \bm{a}_{\text{top}} = \int d\bm{r} ~n_{\text{top}}(\bm{r}) \bm{r}.
\end{eqnarray}

\section{Conclusions\label{Sec:Disc}}

In this work, we have developed a theory of N\'eel-type skyrmions in a weak nonuniform magnetic field. We have introduced and elucidated an \textit{Ansatz} for 
the magnetization, as described by Eq.~\eqref{eq:ansatz_eccentric}. Utilizing this \textit{Ansatz}, we have detailed the calculation of the skyrmion free energy accounting for an external field up to both the first, Eq.~\eqref{eq:tot_expand1_explicit}, and second, Eq.~\eqref{eq:F2-final}, order approximations in $\gamma \ll 1$, where $\gamma$ is the dimensionless external field strength, Eq.~\eqref{eq:gamma-gen-def}. This formulation presents a significantly more tractable form of the free energy minimization problem. Typically, such a minimization problem corresponds to solving a two-dimensional second-order vector differential equation. However, our \textit{Ansatz} simplifies it to a variational problem for three parameters --- $R$, $\delta$, and $\bm{a}$ --- which is computationally more efficient.

Applying the developed methodology, we have examined the behavior of a skyrmion in the stray field of a Pearl vortex. By defining the field strength with the dimensionless parameter $\gamma$, Eq.~\eqref{eq:gamma-def}, we have adapted
the results from the previous sections to this context. Our approach has enabled the identification of numerous metastable skyrmion states within the vortex field. We have demonstrated the potential for different configurations --- both coaxial and displaced from the vortex --- to coexist for certain values of the vortex effective strength~$\gamma$ and DMI parameter~$\epsilon$, Eq.~\eqref{eq:epsilon-def}, as illustrated in the complete phase diagram, Fig.~\ref{fig:phasediag}. Notably, our \textit{Ansatz} facilitates the efficient identification of metastable system minima, which would require specific conditions for detection via numerical solutions in micromagnetic modeling.

Finally, we have sought to compare the outcomes of our analytical calculations with those from micromagnetic simulations. The comparison of parameters $R$ and $a$, as depicted in Fig.~\ref{fig:aRvsgamma}, underscores a remarkable correspondence between the two methodologies, particularly regarding the skyrmion size. Discrepancies in $a$ can be attributed to the flatness of the free energy function near $a_{\rm min}$, which, in turn, influences the sensitivity of the micromagnetic simulation outcomes to the algorithm's tolerance and the system's mesh size. Despite these considerations, the concordance between theoretical predictions and numerical simulations for $a$ is remarkably robust.

The findings presented in this paper can be expanded in several directions. On the one hand, it would be intriguing to explore other geometries, akin to those discussed in 
\cite{hess2023josephson},
or configurations of an inhomogeneous magnetic field, such as those in a vicinity of a planar defect, as proposed by \cite{plastovets2023delocalization},
or near a domain wall \cite{PhysRevB.101.214432}.
Another avenue of research is the investigation of multiple skyrmions in the field of a single vortex or a lattice of vortices. The dynamics of skyrmions in an external inhomogeneous field also presents a significant interest of study, where the magnetization evolves in time, $\bm{m}(\bm{r}, t)$. Lastly, within the context of a skyrmion-vortex pair, examining the emergence and stability of Majorana bound states, with consideration for alterations in the skyrmion profile, is particularly compelling.

\begin{acknowledgements}
The authors are grateful to M.~Shustin, O.~Tretiakov, and P.~Vorobyev for useful discussions.
The work was funded in part by the Russian Science Foundation under the grant No. 21-42-04410. The work of S.S.A. and I.S.B. was funded in part by Basic research program of HSE. We
acknowledge the computing time provided to us at computer facilities at Landau Institute. E.S.A acknowledges support by the Deutsche Forschungsgemeinschaft (DFG, German Research Foundation) within Project-ID 314695032 – SFB 1277 (project A03 and IRTG).
\end{acknowledgements}

\appendix

\section{Euler-Lagrange equations for different skyrmion-vortex configurations\label{app:ELE}}

In this appendix we provide additional details on the derivation of the Euler-Lagrange equations corresponding to a free skyrmion and a skyrmion coaxial to a Pearl vortex.

\subsection{Free skyrmion}

Substituting $\bm{m}_{\rm Sk}$ from Eq. \eqref{eq:m_free} into Eq. \eqref{eq:MagFe}, we find
\begin{eqnarray}
&&\mathcal{F}_{\text{magn}}[\bm{m}_{\rm Sk}] 
= 2\pi d_F \int\limits_0^\infty dr \, r \bigg\{ A \Big[(\partial_r\theta)^2+\frac{\sin^2\theta}{r^2}\Big]
\notag \\
&&\quad 
+D\eta \Big[\partial_r\theta +\frac{\sin2\theta}{2r}\Big]
+ K \sin^2\theta\bigg \} .
\label{eq:FSk0}
\end{eqnarray}

Minimizing $\mathcal{F}_{\text{magn}}[\bm{m}_{\rm Sk}]$ by $\theta(r)$ we derive Eq.~\eqref{eq:ELE_theta_0}.

\subsection{Coaxial skyrmion and vortex}

Substituting  $\bm{m}_{\rm Sk}$ from Eq. \eqref{eq:m_coax} into Eq. \eqref{eq:F_tot}, we find
\begin{eqnarray}
&&\mathcal{F}_{\text{tot}}[\bm{m}_{\rm Sk}] = 2\pi d_F \int\limits_0^\infty dr \, r \Biggl\{ A \Bigl[\theta^{\prime 2}(r)+\frac{\sin^2\theta(r)}{r^2}\Bigr ]
\notag \\
&&\qquad
{}+D\Bigl [\theta^\prime(r) +\frac{\sin2\theta(r)}{2r}\Bigr ]
+ K \sin^2\theta(r)
\notag\\
&& \quad
{}- M_s B_0  [b_z(r)\cos\theta(r)+b_r(r)\sin\theta(r)]\Biggr \} ,
\label{eq:FcSk}
\end{eqnarray}
where $b_{r}(r)$ and $b_{z}(r)$ are the components of the normalized centrally symmetric magnetic field~$\bm{B}$, see Eq.~\eqref{eq:b-symmetr}.

Minimizing $\mathcal{F}_{\text{tot}}[\bm{m}_{\rm Sk}]$ we derive Eq.~\eqref{eq:ELE_theta_coax}.

\section{Magnetization \textit{Ansatz} in local coordinates\label{app:LocalCoord}}

Here we rewrite the developed \textit{Ansatz}, Eq.~\eqref{eq:ansatz_eccentric}, in the special form in local coordinates, which can be useful for some theoretical frameworks. 

We introduce the local orthogonal frame basing on the leading term~$\bar{\bm{m}}_{\rm Sk}$ of the skyrmion magnetization,
\begin{eqnarray}
    \bm{e}_1=\bm{e}_\phi\times\bar{\bm{m}}_{\rm Sk},
    \quad
    \bm{e}_2=\bm{e}_\phi,
    \quad
    \bm{e}_3=\bar{\bm{m}}_{\rm Sk}.
\end{eqnarray}
Then the magnetization \textit{Ansatz}~${\bm{m}}_{\rm Sk}$ defined by Eq.~\eqref{eq:ansatz_eccentric} can be presented as
\begin{eqnarray}
    {\bm{m}}_{\rm Sk}=\bm{e}_+\psi+\bm{e}_-\psi^*+\bm{e}_3\sqrt{1-2|\psi|^2},
\end{eqnarray}
where $\bm{e}_\pm= (\bm{e}_1\pm i\bm{e}_2)/\sqrt{2}$, and the dimensionless complex field $\psi$ is related to $\bm{\mu}^{\bm{a}}$, see Eq.~\eqref{eq:omega2},
\begin{equation}
    \psi=(\mu^{\bm{a}}_r\cos\bar{\theta}-i\mu^{\bm{a}}_\phi)/\sqrt{2}.
\end{equation}

\section{Total free energy in the second order approximation in $\gamma\ll1$\label{app:F2}}

In this appendix we explain why this approximation affects the dependence of the distance~$a$ on $\gamma$, provide additional details on the derivation of the free energy in the second order approximation by the small vortex effective strength, $\gamma\ll1$, and study the behavior near $\gamma_{\rm cr}^{+}$.

\subsection{``Toy model'' of minimization problem with small parameter $\gamma$\label{app:F2-toy1}}

Here we consider a simple problem of minimization of the function $F(p,a;\gamma)$ by two arguments $p$ and $a$, where $\gamma$ is a small parameter,
\begin{equation}
	\label{eq:Fpa-min}
	F(p,a;\gamma)\approx F_0(p)+\gamma F_1(p,a)+\gamma^2 F_2(p,a)\to \min.
\end{equation}
The values of arguments, where the minimum of the function $F(p,a;\gamma)$ is achieved, can be found in the form of a series by $\gamma\ll1$,
\begin{equation}
	\label{eq:pa_min}
	p_{\rm min}=p_0+\gamma p_1, 
\quad
	a_{\rm min}=a_0+\gamma a_1.
\end{equation}

When limiting a problem~\eqref{eq:Fpa-min} to the 
zeroth order approximation in $\gamma$,
\begin{equation}
	\label{eq:Fpa-min0}
	F(p,a;\gamma)\approx F_0(p)\to \min ,
\end{equation}
we have lost the dependence on $a$ and, minimizing, can find only $p_{\rm min}=p_0$ from condition $\partial_p F_0(p_0)=0$.

In order to find value $a_{\rm min}$, we extend the problem to the first approximation in $\gamma$,
\begin{equation}
	\label{eq:Fpa-min1}
	F(p,a;\gamma)\approx F_0(p)+\gamma F_1(p,a)\to \min.
\end{equation}
Applying here Eqs.~\eqref{eq:pa_min} and minimizing, one can find $a_{\rm min}=a_0$ from condition $\partial_a F_1(p_0,a_0)=0$, and
\begin{equation}\label{eq:p0-p1}
	{p_{\rm min}=p_0+\gamma p_1},
	\qquad
	p_1=\dfrac{\partial_p F_1(p_0,a_0)}{\partial_p^2 F_0(p_0)}.
\end{equation}

Now to determine dependence on $\gamma$ in $a_{\rm min}$, we should extend the problem to the second approximation in $\gamma$, as in Eq.~\eqref{eq:Fpa-min}. Minimizing we find 
\begin{equation}
	a_{\rm min}=a_0+\gamma a_1,
	\quad
	a_1=-\frac{p_{1} \partial_{p,a}^2F_{1}+\partial_{a}F_{2}}{\partial_{a}^2F_{1}}\Big|_{{\scriptsize p=p_0}\atop{\scriptsize a=a_0}}.
 \label{eq:a1}
\end{equation}
So, we can see that to determine the dependence $a_{\rm min}$ on $\gamma$ one should expand $F(p,a;\gamma)$ to the second order in $\gamma$ at least. 

Moreover, if $p_{\rm min}$ and $a_{\rm min}$ depend on $\gamma$ in more complicated manner (see the next subsection) than linear, but variate from $p_{0}$ and $a_{0}$ weakly, 
\begin{equation}
    |p_{\rm min}-p_0|\ll p_0,
    \quad
    |a_{\rm min}-a_0|\ll a_0,
\end{equation}
we can use Eq.~\eqref{eq:a1} rewritten in the following form
\begin{equation}
	a_0-a_{\rm min}(\gamma)=\frac{[p_{\rm min}(\gamma)-p_0] \partial_{p,a}^2F_{1}+\gamma \partial_{a}F_{2}}{\partial_{a}^2F_{1}}\Big|_{{\scriptsize p=p_0}\atop{\scriptsize a=a_0}}.
 \label{eq:a1-general}
\end{equation}
Indeed, the curves in Fig.~\ref{fig:aRvsgamma} are in agreement with the last expression for $\gamma\ll1$, i.e., partly for $\epsilon\lesssim0.4$ and for whole curves for $\epsilon\gtrsim0.4$.

\subsection{Calculation of the $\mathcal{F}^{(2)}$\label{app:F2-calc}}

When deriving expression for $\mathcal{F}^{(2)}[\bar{\bm{m}},\tilde{\bm{m}},\bm{a}]$, we keep the only terms of order of $\gamma^2$ that are presented in Eq.~\eqref{eq:F2}. The other possible contributions appear to be of the order of $\gamma^3$ or smaller. In particular, in Eq.~\eqref{eq:F2-spare1} the contribution $\mathcal{F}_{\text{magn}}^{(1)}[\bar{\bm{m}}_{\rm Sk},\gamma^2\bm{\lambda}(1-\bm{\lambda}\cdot\bar{\bm{m}})]$ is calculated and proven to be of order of $\gamma^3$. The other contributions that can be neglected in Eq.~\eqref{eq:F2} are
\begin{eqnarray}
	\mathcal{F}_{\text{magn}}^{(1)}[\gamma\tilde{\bm{m}}_{\rm Sk},\gamma^2\{\ldots\}] 
	\text{ and }	
	\mathcal{F}_{\text{tot}}'[\gamma^2\{\ldots\},\bm{B}^{\bm{a}}],
\end{eqnarray}
where $\{\ldots\}$ means the whole expression in the curly brackets in Eq.~\eqref{eq:ansatz_eccentric2}.

\subsection{Dependencies $a(\gamma)$ and $R(\gamma)$ near $\gamma\approx\gamma_{\rm cr}^+$\label{app:F2-toy2}}

As we can see from Fig.~\ref{fig:F(a)} there is a critical value $\gamma_{\rm cr}^+$, when maximum and minimum of $\mathfrak{F}(a)$ merge and turn into the inflection point, $a=a_{\rm in}$. For $\gamma>\gamma_{\rm cr}^+$ the only minimum is at $a=0$. Considering $\mathfrak{F}(a)$ for $\gamma\approx\gamma_{\rm cr}^+$ and $a\approx a_{\rm in}$, we can conclude that it should look like as
\begin{equation}
    \mathfrak{F}(a)\approx\mathfrak{F}(a_{\rm in})+\alpha(\gamma_{\rm cr}^+-\gamma)(a-a_{\rm in})+\beta(a-a_{\rm in})^3/3,
    \label{eq:C3}
\end{equation}
where $\mathfrak{F}(a_{\rm in})$, $\alpha$, and $\beta$ some constants. Minimizing 
the function \eqref{eq:C3} over $a$, we find that
\begin{equation}
\label{eq:amin-gammacr}
    a_{\rm min/max}(\gamma)=a_{\rm in}\pm\sqrt{\gamma_{\rm cr}^+-\gamma}\sqrt{\alpha/\beta}.
\end{equation}

Based on Eq.~\eqref{eq:a1-general} we can conclude that $R_{\rm min}(\gamma)$ should have the same behavior $R_{\rm in}-R_{\rm min}(\gamma)\propto\sqrt{\gamma_{\rm cr}^+-\gamma}$, where $R_{\rm in}$ is a radius of a metastable skyrmion state that corresponds to $a=a_{\rm in}$ and $\gamma=\gamma_{\rm cr}^+$.

\section{Stray magnetic field of a Pearl vortex\label{app:vortex}}

\subsection{Arbitrary thickness of a superconducting film}

The stray magnetic field of the vortex produced outside of the superconducting film of thickness~$d_S$ is given by~\cite{Carneiro2000}
\begin{equation}
B_{{V},{r}/z}(r,z>0)=-\frac{\phi_0}{2\pi}\int_0^{\infty}\dfrac{dq\,qJ_{1/0}(qr)}{F(q)}e^{-qz}, 
\label{eq:SM_B}
\end{equation}
where 
\begin{equation}
F(q)=\lambda_L^2\tau \frac{(q+\tau)^2e^{\tau d_S}-(q-\tau)^2e^{-\tau d_S}}{(q+\tau)e^{\tau d_S}+(q-\tau)e^{-\tau d_S}-2q},
\label{eq:SM_Fq}
\end{equation}
and $\tau=\sqrt{q^2+\lambda_L^{-2}}$.

\subsection{Thin superconducting film}

In the limits $d_S\to0$ and $\lambda_L\to0$, while the Pearl length $\lambda=\lambda_L^2/d_S$ is kept constant~\cite{Pearl1964}, we arrive to 
\begin{equation}
\label{eq:SM_Fq_Pearl}
F(q)=1+2 q\lambda.
\end{equation}
Equations~\eqref{eq:SM_B} with the latter expression applied to define the same stray field that Eq.~\eqref{eq:vortex:B}.

Therefore, normalizing the stray field at $z\to+0$ by means of Eq.~\eqref{eq:vortex:B:thin-F}, we get
\begin{gather}
b_{r/z}(r)=-2\ell_{w}\intl_{0}^{\infty} \frac{dq\,qJ_{1/0}(qr) }{\lambda^{-1}+2q}.
\label{eq:vortex:h}
\end{gather}

\subsection{Large Pearl length}

In this paper we assume that the skyrmion radius $|R|$ as well as domain wall width~$\ell_w$ are small in comparison to Pearl length, see Eq.~\eqref{eq:assumptions}. Under this assumption one can treat Eq.~\eqref{eq:vortex:h} for $r\ll\lambda$ only, neglect $\lambda^{-1}$ in the denominator, and arrive immediately to Eq.~\eqref{eq:vortex:B:b0_b1}. 

\subsection{``No-skyrmion'' configuration in the stray field of a Pearl vortex\label{app:vortex-only}}

The magnetization angle described by Eq.~\eqref{eq:thetagamma-ODE} with boundary conditions~\eqref{eq:bound_theta_b} can be calculated analytically for the vortex stray field. Taking $b_r(r)$ from Eq.~\eqref{eq:SM_B}, we get
\begin{equation}
\theta_{\bm{b}}(r)=-2\ell_{w}\lambda\intl_{0}^{\infty} \frac{dq\,qJ_{1}(qr) }{F(q)[1+(\ell_{w}q)^2]}.
\end{equation}

Assuming thin superconducting layer and using Eq.~\eqref{eq:SM_Fq_Pearl}, we can write the latter expression as
\begin{equation}
\theta_{\bm{b}}(r)=-2\ell_{w}\intl_{0}^{\infty} \frac{dq\,qJ_{1}(qr) }{(\lambda^{-1}+2q)[1+(\ell_{w}q)^2]}.
\end{equation}

Finally, assuming $r\ll\lambda$, we arrive at Eq.~\eqref{eq:thetagamma-approx}.

\subsection{Averaged eccentric vortex\label{app:averaged_vortex}}

Considering the integral $\mathcal{F}_{Z}[\bar{\bm{m}}_{\rm Sk},\bm{B}_V^{\bm{a}}]$ in polar coordinates with origin in $\bm{a}$ we can write it in the form of Eq.~\eqref{eq:F-Sk-V-0}, 
\begin{gather}
\mathcal{F}_{Z}[\bar{\bm{m}}_{\rm Sk},\bm{B}_V^{\bm{a}}]  = - d_F  \int d^2 \bm{r} M_s \bar{\bm{m}}_{\rm Sk}\cdot \bm{B}_{V}
^{\bm{a}}|_{z=+0}.
\label{eq:F-Sk-V-a}
\end{gather}
Since $\bar{\bm{m}}_{\rm Sk}$ is radially symmetric, we can integrate over polar angle (average) only the stray fields, cf. Eq.~\eqref{eq:B-aver},
\begin{equation}
\bm{B}_{V}^{\bm a} \to \bar{\bm{B}}_{V}^{\bm a}=
-\dfrac{\phi_0}{4\pi \ell_{w}\lambda}
\big[
\bar{b}_r^{\bm a}(r)\bm{e}_{r}+\bar{b}_z^{\bm a}(r){\bm e}_z
\big].
\label{eq:BV-aver}
\end{equation}
Substituting into Eq.~\eqref{eq:F-Sk-V-a} the expression for~$\bar{\bm{m}}_{\rm Sk}$ via $\bar{\theta}(r)$, we arrive at the last line of Eq.~\eqref{eq:tot_expand1_explicit}.

The exact expressions for functions $\bar{b}_{r/z}^{\bm{a}}(r)$ can be obtained from Eq.~\eqref{eq:SM_B} by definition of Eq.~\eqref{eq:B-aver}, 
\begin{equation}
\bar{b}_{r/z}^{\bm a}(r)= - 2\ell_w\lambda\int\limits_0^\infty \frac{dq\,q J_0(qa) J_{1/0}(qr)}{F(q)},
\label{eq:vortex:field1}
\end{equation}
or for the thin films,
\begin{equation}
\bar{b}_{r/z}^{\bm a}(r)= - 2\ell_w\int\limits_0^\infty \frac{dq\,q J_0(qa) J_{1/0}(qr)}{\lambda^{-1}+2q}
\label{eq:vortex:field2}
\end{equation}

Finally, assuming $r\ll\lambda$, we arrive at Eqs.~\eqref{eq:vortex:field0}.

Applying $\bar{b}_{r}^{\bm a}(r)$ from Eq.~\eqref{eq:vortex:field0} to Eq.~\eqref{eq:thetagamma-ODE}, we can find $\theta_{\bar{\bm{b}}^{\bm a}}(r)$ in the form of Eq.~\eqref{eq:thetaAgamma-approx}.

\section{Details of micromagnetic simulations\label{app:mmm}}

In our micromagnetic simulations, we employed the discretized lattice version of the following Hamiltonian:
\begin{equation}
    H = (\nabla \bm{m})^2 - \epsilon(\bm{m} \cdot \nabla m_z - m_z \nabla \cdot \bm{m}) - m_z^2 - 2 \gamma \bm{m} \cdot \bm{b}^{\bm{a}}.
\end{equation}
Assuming that the gradients are small and that $\bm{m}(\bm{r})$ is a smooth and slow function of distance, the derivatives in the continuous-limit Hamiltonian can be effectively mapped onto a finite-mesh lattice provided that $|dL \cdot \partial_{i} m^j| \ll 1$ for $i = x,y$ and $j = x,y,z$. Next, we numerically minimize the energy with respect to the unknown field $\bm{m}(\bm{r})$.

In our analysis, we used square samples of size $2L \times 2L$ with a square cell dimension of $dL \times dL$. We selected $L = 10.05 \ell_w$ and $dL = 0.015 \ell_w$ to accommodate a range of skyrmion radii ($R \sim 0.2 - 1.5 \ell_w$) and displacements between skyrmions and vortices ($a_{\rm min} \sim 1.5-3 \ell_w$). It is important to note that this wide spread of parameters imposes significant limitations on our modeling capabilities. For example, at $\epsilon = 0.25$, we are constrained to $\gamma \geqslant 0.3$. This restriction is due to the fact that for smaller values of $\gamma$, the skyrmion radius decreases, the depth of the free energy minimum becomes less pronounced, and the distance $a_{\rm min}$ increases, all of which require extensive computational resources.

To determine the typical parameters $R$, $\delta$, and $\bm{a}$, we utilized Eq.~\eqref{eq:ansatz_eccentric} as a fitting model for our relaxed magnetization profiles. The corresponding parameters $R$ and $a$ are depicted in Fig.~\ref{fig:aRvsgamma}. Deformations from the cylindrical shape of the skyrmion result in the insufficiency of the original domain wall \textit{Ansatz}, as given by Eq.~\eqref{eq:ansatz_free}, to accurately represent the altered profile shape. Conversely, our proposed \textit{Ansatz}, Eq.~\eqref{eq:ansatz_eccentric}, accounts for these changes and yields consistent profiles, meaning that the simulation results are well approximated by this model.

\bibliography{bib-skyrmion}

\end{document}